\documentclass[10pt,aps,prl,twocolumn,superscriptaddress,amsmath,amssymb,nofootinbib,floatfix]{revtex4-2}
\usepackage[normalem]{ulem}
\usepackage{graphicx,bm,mathtools,xcolor}
\usepackage{orcidlink}
\usepackage{tikz}
\usetikzlibrary{arrows.meta,calc,positioning,fit,backgrounds}
\usepackage{hyperref}
\hypersetup{colorlinks=true,allcolors=orange!100!black}

\newcommand{\ms}{\mathfrak{m}_2}            
\newcommand{\Mf}{\mathcal{M}}               
\newcommand{\wt}{w}                          
\newcommand{\avg}[1]{\overline{#1}}

\begin{document}

\title{Quasiparticle Theory of Nonlocal Magic in Monitored Fermionic Gaussian Systems}
\author{Akash Kumar Heera~\orcidlink{0009-0009-5388-3162}}
\affiliation{International School for Advanced Studies (SISSA), Via Bonomea 265, 34136 Trieste, Italy}
\author{Alessandro Silva~\orcidlink{0000-0002-5668-7134}}
\affiliation{International School for Advanced Studies (SISSA), Via Bonomea 265, 34136 Trieste, Italy}
\author{Mario Collura~\orcidlink{0000-0003-2615-8140}}
\affiliation{International School for Advanced Studies (SISSA), Via Bonomea 265, 34136 Trieste, Italy}
\affiliation{INFN, Sezione di Trieste, Via Valerio 2, 34127 Trieste, Italy}
\date{\today}

\begin{abstract}
We propose a quasiparticle theory for the dynamics of fermionic nonlocal magic in monitored free fermions. From a stabilizer product state, correlations propagate as quasiparticle multiplets, which quantum jumps collapse and renew into pairs whose nonlocal magic is set by the local filling. Without free parameters, the theory quantitatively reproduces quantum-jump simulations of two number-conserving chains after homogeneous, two-reservoir, and domain-wall quenches, and predicts that arbitrarily weak monitoring turns volume-law nonlocal magic into an area law.
\end{abstract}

\maketitle

\paragraph{Introduction.---}

Entanglement alone does not capture the complexity of a quantum state. Stabilizer states, for instance, can be highly entangled and yet are efficiently simulable on a classical computer~\cite{Gottesman1998,Aaronson2004}; what promotes them to universal quantum computation is nonstabilizerness, or ``magic''~\cite{Bravyi_2005,Veitch2014,Howard2017}. Magic can be quantified by several monotones, such as the log-Wigner negativity and the relative entropy of magic~\cite{Veitch2014}, the robustness of magic~\cite{Howard2017}, the stabilizer rank and extent~\cite{BravyiSmithSmolin2016,Bravyi2019simulation}, the stabilizer nullity~\cite{Beverland2020}, the thauma~\cite{WangWildeSu2020}, generalized robustness measures~\cite{Seddon2021,Liu2022}, and the stabilizer R\'enyi entropies (SREs)~\cite{Leo21b,LeoneBittel2024,HaugPiroli2023}, which avoid any optimization over stabilizer states. The interplay between the two resources raises a natural question: how much of the magic of a bipartite state is stored in its correlations, rather than in the local bases? This is captured by the nonlocal magic, obtained by minimizing a magic measure over local unitaries~\cite{cao2024,cao25,Qian_2025,Andreadakis2026,Cepollaro2025,Robin2026}, which has recently been measured on a superconducting quantum processor~\cite{Ahmad2025}. When the optimization runs over arbitrary local unitaries, the nonlocal magic is controlled by the entanglement spectrum and grows at most logarithmically with the entanglement entropy~\cite{sierant2026,Sierant2026exact,Huang2026,Torre2026,Liu2026,Franchini2026}.

The situation is different for fermionic Gaussian states. These states are classically simulable as well~\cite{Valiant2002,TerhalDiVincenzo2002,Knill2001,JozsaMiyake2008,Bravyi2005}, and for them non-Gaussianity plays the role of the computational resource~\cite{Hebenstreit2019,Oszmaniec2022,DiasKoenig2024,Sierant2026FMR}. If the optimization is restricted to local \emph{Gaussian} unitaries, the resulting fermionic nonlocal magic (FNLM) admits a closed form in terms of the reduced covariance matrix~\cite{Iannotti2026,collura2026}, see Ref.~\cite{Collura2026quantum} for the SRE of Gaussian states; it can be extensive and, since it relies only on two-point functions, it is accessible through fermionic shadow tomography~\cite{ZhaoRubinMiyake2021}. In the following, ``nonlocal magic'' always refers to this Gaussian-restricted quantity, which upper-bounds the unrestricted one.

After a quantum quench in an integrable system, the spreading of entanglement can be understood in terms of correlated quasiparticles propagating ballistically~\cite{CalabreseCardy2005,Fag2008,AlbaCalabrese2017,AlbaCalabrese2018,Calabrese2020notes}, organized in pairs or, for more general initial states, in multiplets~\cite{Bertini2018,Bastianello_2018,BastianelloCollura2020}. The same picture applies to the FNLM, but with a different weight carried by each quasiparticle~\cite{Iannotti2026,collura2026}, and quasiparticle descriptions have also been developed for the unrestricted nonlocal magic~\cite{Aditya2026}. Continuous monitoring introduces a competing mechanism, since local measurements tend to disentangle the system and can drive measurement-induced transitions~\cite{Skinner2019,Li2018,Li2019,Chan2019,Fisher2023,PotterVasseur2022}. For monitored free fermions, Cao, Tilloy, and De Luca proposed that measurements collapse quasiparticle pairs and renew them in place, an ansatz that quantitatively describes the growth and saturation of entanglement~\cite{Cao2019}. Related ideas have been used to describe the stochastic resetting of non-Hermitian quasiparticles~\cite{TurkeshiResetting2022,TurkeshiResettingErratum2023}, dissipative quasiparticle pictures~\cite{Car21,AlbaCarollo2021} and monitored multiplets~\cite{Car22}, monitored harmonic chains~\cite{Young2024}, and monitored quantum point contacts~\cite{Delmonte2026}. For monitored chains with U(1) symmetry, it is now understood that the stationary entanglement obeys an area law for any measurement rate~\cite{Cao2019,Poboiko2023,Nie25}, even though apparent transitions appear at accessible system sizes~\cite{Alb20c,Fuj20,Buchhold2021,Coppola2022,Ladewig2022,Pic24}. On the magic side, the nonstabilizerness of monitored Gaussian fermions has been studied numerically~\cite{Tir25,wang2025}, and magic transitions have been reported in monitored circuits~\cite{Bejan2024,Niroula2024,Fux2024,TarabungaTirrito2024}. Whether a quasiparticle description can account for the dynamics of nonlocal magic under monitoring, however, remains an open question, all the more so because nonlocal magic and entanglement weigh the same correlations differently, so that the fate of one under measurements does not fix the fate of the other.

\begin{figure}[htbp]
\centering
\includegraphics[width=\linewidth]{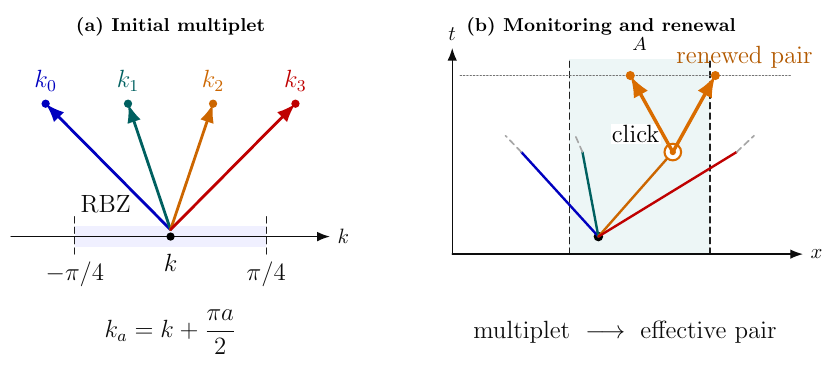}
\caption{Illustration of the quasiparticle-collapse picture.
(a) The period-four product state emits correlated multiplets of four modes $k_a=k+\pi a/2$, with the band index (member index) $a=0,1,2,3$,
which propagate with velocities $v_{k_a}$; the contribution of a multiplet to the nonlocal magic of $A$ depends on how many of its members lie inside $A$.
(b) Quantum jumps collapse each multiplet as a whole at rate $\gamma$ and replace it with a single pair with velocities $\pm v_k$, which carries the nonlocal magic weight $\ms(\rho)$ set by the local filling $\rho$.}
\label{fig:qp-collapse}
\end{figure}

In this work, we propose such a description. Motivated by the collapsed-quasiparticle ansatz of Ref.~\cite{Cao2019}, we view the disentangled initial state as a source of quasiparticle multiplets~\cite{Bertini2018,BastianelloCollura2020}, and consider initial states whose multiplets generate nonlocal magic at leading order as they spread, such as periodic product states. Quantum jumps collapse these multiplets at a rate set by the jump statistics and renew them into pairs, whose nonlocal magic weight is set by the local filling (Fig.~\ref{fig:qp-collapse}). The resulting ansatz has no free parameters and is local, so that it applies to homogeneous and inhomogeneous quenches alike. A striking consequence is that arbitrarily weak monitoring destroys volume-law nonlocal magic: since quasiparticle correlations live for a time $\sim1/\gamma$, nonlocal magic cannot be built beyond the length $\ell^*\sim\bar v/\gamma$, where $\bar v$ is a typical quasiparticle velocity, and the stationary nonlocal magic crosses over from a volume law to an area law. We test these predictions against quantum-jump simulations of two number-conserving chains with different dispersions, in homogeneous quenches from periodic states at several fillings and in two inhomogeneous quenches: a two-reservoir quench, joining two period-four halves with shifted occupation patterns, and a domain-wall quench. We also identify a transient dip of the nonlocal magic, absent in any pair picture and in the entanglement, which provides a direct fingerprint of the multiplet structure. Finally, for the transverse-field Ising chain with physical occupation monitoring, a non-Hermitian version of the unitary quasiparticle theory captures the no-click dynamics (Appendix~\ref{app:noclick}).

\paragraph{Fermionic nonlocal magic.---}
For a pure state $|\phi\rangle$ of $N$ qubits, the second SRE reads~\cite{Leo21b}
\begin{equation}
M_2(\phi)=-\log_2\Big(\frac{1}{2^N}\sum_{P\in\mathcal P_N}\langle\phi|P|\phi\rangle^4\Big),
\label{eq:SRE}
\end{equation}
where $\mathcal P_N=\{I,X,Y,Z\}^{\otimes N}$ is the set of Pauli strings, and fermionic modes are mapped onto qubits through the Jordan--Wigner transformation. For a pure fermionic Gaussian state $|\psi\rangle$ and a bipartition $A\cup B$, we define the FNLM as the SRE minimized over local Gaussian unitaries,
\begin{equation}
\Mf_A(\psi)=\min_{U_A\otimes U_B\ \mathrm{Gaussian}} M_2\big[(U_A\otimes U_B)|\psi\rangle\big].
\label{eq:FNLdef}
\end{equation}
Introducing the Majorana operators $a_{2j-1}=c_j+c_j^\dagger$ and $a_{2j}=i(c_j^\dagger-c_j)$, the reduced covariance matrix $(\Gamma_A)_{mn}=\tfrac{i}{2}\langle[a_m,a_n]\rangle$, with $m,n\in A$, is real and antisymmetric, so that $i\Gamma_A$ has eigenvalues $\pm\nu_j$ with $\nu_j\in[0,1]$~\cite{SuraceTagliacozzo2022}. The minimum in Eq.~\eqref{eq:FNLdef} is given by the closed form~\cite{Iannotti2026,collura2026},
\begin{equation}
\Mf_A=\sum_{j=1}^{\ell}\ms(\nu_j^2),\qquad \ms(x)=-\log_2(1-x+x^2),
\label{eq:magic}
\end{equation}
where $\ell=|A|$, and each term is nothing but the SRE of the two-mode Gaussian normal form associated with $\nu_j$~\cite{SM}. The entanglement entropy is a function of the same $\nu_j$~\cite{Peschel2003,Vidal2003,PeschelEisler2009}; however, the two quantities weight the spectrum differently, since $\nu_j=0$ maximizes the entanglement but gives no contribution to the FNLM. For number-conserving states one has $\nu_j=|1-2\lambda_j|$, where $\lambda_j$ are the eigenvalues of $(C_A)_{ij}=\langle c_i^\dagger c_j\rangle$, so that each eigenvalue contributes
\begin{equation}
\wt(\lambda)=\ms\big[(1-2\lambda)^2\big],
\label{eq:weight}
\end{equation}
which vanishes at $\lambda=0,\tfrac12,1$ and is maximal, $\log_2(4/3)$, at $\lambda=(2\mp\sqrt2)/4$. In the following, we denote by $\Mf_\ell$ the FNLM of an interval of $\ell$ sites. Under monitoring, we evaluate Eq.~\eqref{eq:magic} on each conditional pure state and then average over quantum trajectories, which we denote by $\avg{\Mf}_\ell$.

\paragraph{Models.---}
To test the theory, we consider two chains of the form $H=\sum_k\epsilon_k d_k^\dagger d_k$, monitored through the jump operators $L_j=\sqrt\gamma\,n_j$, with $n_j=d_j^\dagger d_j$, within the quantum-jump unraveling~\cite{DalibardCastinMolmer1992,Gardiner1992,PlenioKnight1998,Daley2014,WisemanMilburn2009}. (i)~\emph{Hopping chain.}---Here $d_j$ are the physical fermions on the physical lattice sites, with nearest-neighbour hopping and $\epsilon_k=-J\cos k$. The jumps measure the site occupations, and $H$ conserves the particle number $N=\sum_jn_j$. (ii)~\emph{Ising-dispersion chain.}---Here $d_j=L^{-1/2}\sum_ke^{ikj}\eta_k$ are the Wannier orbitals of the Bogoliubov quasiparticles $\eta_k$ of the transverse-field Ising chain, with $\epsilon_k=2\sqrt{(h-J\cos k)^2+J^2\sin^2k}$. The jumps measure the occupations of these orbitals, and $H$ conserves the number of Bogoliubov quasiparticles $N=\sum_jn_j=\sum_k\eta_k^\dagger\eta_k$. Since these modes are nonlocal in the physical spins, this is not the physically monitored Ising chain, which we discuss in Appendix~\ref{app:noclick}; it changes the velocity distribution at fixed measurement structure.

In both chains, $n_j^2=n_j$ and the conservation of $N$ give $H_{\rm eff}=H-i\gamma N/2$: the evolution between jumps is unitary, jumps occur at the total rate $\gamma N$, and conditional states remain Slater determinants. We use rings for homogeneous quenches and open chains for inhomogeneous ones~\cite{SM}.

\begin{figure*}[t!]
\centering
\includegraphics[width=\textwidth]{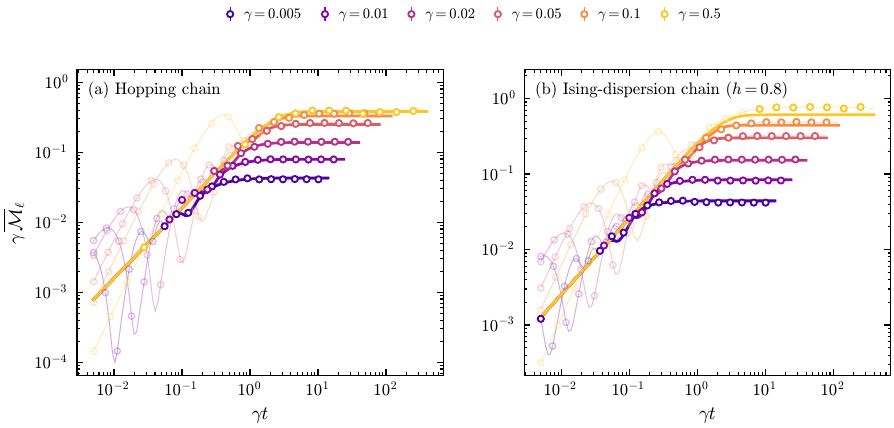}
\caption{Rescaled trajectory-averaged nonlocal magic $\gamma\avg{\Mf}_\ell$ as a function of $\gamma t$ after a quench from $|1000\,1000\cdots\rangle$ in (a) the hopping chain and (b) the Ising-dispersion chain ($h=0.8$), for $\gamma=0.005$, $0.01$, $0.02$, $0.05$, $0.1$, $0.5$ (from dark purple to yellow), with $L=512$, $\ell=32$, $J=1$, and periodic boundary conditions. Open symbols denote quantum-jump averages with standard errors over $N_{\rm traj}=200$ trajectories, and thin lines join all data points; solid lines show Eq.~\eqref{eq:hybrid}, with no free parameters. Faded symbols mark early-time oscillations, $Jt<10$.}
\label{fig:four_models_magic}
\end{figure*}

\paragraph{Initial state and multiplets.---}
The choice of the initial state is crucial. Since $\wt(\lambda)$ vanishes at $\lambda=0,\frac12,1$, the initial correlations contribute to the FNLM at leading order only if they generate other eigenvalues. We therefore consider disentangled initial states whose correlations are carried by quasiparticle multiplets that generate FNLM at leading order once their members are shared between $A$ and its complement. The simplest examples are the periodic product states $|1\,\underbrace{0\cdots0}_{p-1}\,1\,\underbrace{0\cdots0}_{p-1}\,\cdots\rangle$, with one particle per unit cell of length $p$ and filling $\rho=1/p$. Such a state is invariant under translations by $p$ sites, so it can only correlate momenta that differ by multiples of $2\pi/p$. Explicitly, rewriting the occupied sites in momentum space gives, up to a global phase,
\begin{equation}
|1\,\underbrace{0\cdots0}_{p-1}\,1\,\underbrace{0\cdots0}_{p-1}\,\cdots\rangle=\prod_{k\in{\rm RBZ}}\frac{1}{\sqrt p}\sum_{a=0}^{p-1}d^\dagger_{k_a}|0\rangle,
\label{eq:multiplet_state}
\end{equation}
where $k_a=k+2\pi a/p$ and the reduced Brillouin zone (RBZ) is $[-\pi/p,\pi/p)$~\cite{Bertini2018,BastianelloCollura2020,SM}. Each factor describes one fermion is a coherent superposition of the $p$ momenta $k_a$, which form a multiplet with correlation matrix $Q_{ab}=1/p$. When $m$ of its members lie in $A$, the multiplet contributes a single eigenvalue $m/p$, with weight $\mu_p(m)=\wt(m/p)$; more generally, a multiplet carrying one fermion with amplitudes $\psi_a$ on its members contributes the eigenvalue $\sum_{a\in A}|\psi_a|^2$~\cite{SM}. The unitary prediction reads
\begin{equation}
\Mf^{\rm mult}_\ell(t)=\int dx\int_{\rm RBZ}\frac{dk}{2\pi}\;\mu_p[m(k,x,t)],
\label{eq:multiplet}
\end{equation}
where $m(k,x,t)$ counts the members of the multiplet emitted at $x$ that lie in $A$ at time $t$; at long times, $\Mf^{\rm mult}_\ell\to\ell\,\wt(\rho)$. The multiplets contribute at leading order for $p\geq3$. For the N\'eel state, $p=2$, instead, the only multiplets shared between $A$ and its complement have $m=1$, i.e., eigenvalue $\frac12$ and zero weight~\cite{SM}: its FNLM is subleading and grows logarithmically~\cite{collura2026}. Since the multiplets emitted at $x$ are fixed by the initial state within a distance of order $p$ of $x$~\cite{Bertini_2018,Alba2018}, Eq.~\eqref{eq:multiplet} also applies to locally periodic states, with $p$, the RBZ, and $\mu_p$ taken at the emission point, and with no emission from empty regions~\cite{SM}. For $p\geq3$, Eq.~\eqref{eq:multiplet} reproduces the exact Gaussian dynamics, whereas the conventional pair approximation fails~\cite{SM}. Our examples use $p=4$, for which $\mu_4(1)=\mu_4(3)=\ms(1/4)=\log_2(16/13)$ and $\mu_4(2)=0$.

A direct signature of the multiplets is a transient dip of the FNLM, absent in any pair picture: when the fastest members leave $A$, multiplets with $m=3$ members inside, of weight $\mu_4(1)$, turn into multiplets with $m=2$, of zero weight. Further details are given in Ref.~\cite{SM}. The dip survives weak monitoring (Fig.~\ref{fig:four_models_magic}).

\paragraph{Collapse ansatz.---}
We now include the measurements. As in the unitary case, we assume a disentangled initial state whose quasiparticle multiplets generate nonlocal magic at leading order as they spread, with one particle per multiplet and weights $\mu_p(m)=\wt(m/p)$, and denote by $\Mf^{\rm mult}_\ell(t)$ their unitary contribution [Eq.~\eqref{eq:multiplet}]. Motivated by Ref.~\cite{Cao2019}, the ansatz postulates that: (i) during each time interval $dt$, a multiplet is collapsed as a whole by a quantum jump with probability $\gamma\,dt$, which destroys the correlations among all its members; (ii) each collapse creates a single new pair of quasiparticles at the position of the jump, with momenta $\pm k$ and $k$ uniformly distributed in $[-\pi,\pi)$, and renewed pairs are themselves collapsed and renewed in the same way; and (iii) a pair emitted at position $x$ and time $s$ carries the nonlocal magic weight $\ms[\rho(x,s)]$, set by the local filling $\rho(x,s)$ and independent of $k$, whereas the multiplets of the initial state keep their weights $\mu_p(m)$. Multiplets and pairs otherwise evolve independently. Collecting the surviving multiplets and the pairs emitted at time $t-a$, we obtain
\begin{align}
\avg{\Mf}_\ell(t)&\simeq{}e^{-\gamma t}\,\Mf^{\rm mult}_\ell(t)\nonumber\\
&+\gamma\!\!\int_0^t\!da\,e^{-\gamma a}\!\!\int_{-\pi}^{\pi}\!\frac{dk}{2\pi}\int\!dx\,\chi_k(x,a)\,\ms[\rho(x,t-a)],
\label{eq:renewal}
\end{align}
where $\chi_k(x,a)=1$ if the member at $x+v_ka$ of a pair emitted at $x$ lies in $A$ while its partner at $x-v_ka$ does not, and $\chi_k=0$ otherwise, with $v_k=\partial_k\epsilon_k$. The first term accounts for the initial multiplets that have not been collapsed yet, while the second collects the pairs emitted at time $t-a$, which have propagated for a time $a$.

The ansatz is motivated by the jump update: a jump at site $j$ replaces the occupied orbital $C|j\rangle/\sqrt{\langle n_j\rangle}$ with the orbital $|j\rangle$, which is localized at a single site and hence contains all momenta with equal weight, while the rest of the occupied subspace is unchanged~\cite{SM}. Since jumps occur at total rate $\gamma N$, each of the $N$ occupied orbitals is renewed at the average rate $\gamma$. With one particle per unit cell, each multiplet is exactly one occupied orbital, and we take it to survive as a whole with probability $e^{-\gamma t}$. We verify directly that the initial multiplet correlations decay at the rate $\gamma$~\cite{SM}. These arguments motivate, but do not derive, the postulates; we test the ansatz numerically.

The local filling follows from the averaged dynamics. The trajectory-averaged correlation matrix obeys exactly the closed equation $\partial_t\avg C=-i[h,\avg C]-\gamma[\avg C-\mathrm{diag}(\avg C)]$~\cite{SM}, which is nothing but the dephasing master equation~\cite{Eisler2011,Znidaric2010,BernardJin2019}. To leading order in gradients, its Wigner function $n(x,k,t)$ then obeys
\begin{equation}
\partial_t n=-v_k\partial_x n-\gamma\big[n-\rho(x,t)\big],\qquad \rho=\int\frac{dk}{2\pi}\,n.
\label{eq:kinetic}
\end{equation}
Note that this equation fixes only the filling, not the weight: the FNLM is nonlinear in the correlation matrix, so its trajectory average is not determined by $\avg C$, and the form $\ms(\rho)$ is the content of postulate (iii). For a homogeneous quench from a product state in the occupation basis, the Wigner function is $k$-independent, $n(x,k,0)=\langle n_x\rangle$, which after coarse-graining over a unit cell gives $n=\rho=1/p$. This is a stationary solution of Eq.~\eqref{eq:kinetic}, so that $\rho(x,t)=\rho$ at all times. The $x$-integral in Eq.~\eqref{eq:renewal} then gives $\min(2|v_k|a,\ell)$, and
\begin{align}
\avg{\Mf}_\ell(t)\simeq{}&e^{-\gamma t}\,\Mf^{\rm mult}_\ell(t)\nonumber\\
&+\gamma\,\ms(\rho)\int_0^t\!da\,e^{-\gamma a}\int_{-\pi}^{\pi}\!\frac{dk}{2\pi}\,\min(2|v_k|a,\ell),
\label{eq:hybrid}
\end{align}
which contains no free parameters: the multiplet term is fixed by the initial state, and the renewal weight by its filling. As for the quasiparticle picture without monitoring~\cite{Fag2008,Bertini2018}, we expect Eq.~\eqref{eq:hybrid} to hold for $p\ll\ell\ll L$, with $\ell^*\gg p$. For an inhomogeneous initial filling, $\rho(x,t)$ instead follows from integrating Eq.~\eqref{eq:kinetic} from the coarse-grained initial occupation, and Eq.~\eqref{eq:renewal} is evaluated directly; the prediction remains free of parameters.

\begin{figure}[t]
\centering
\includegraphics[width=\linewidth]{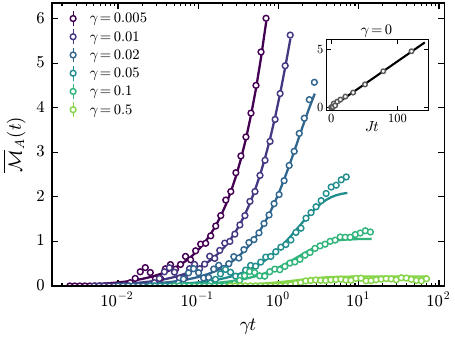}
\caption{Trajectory-averaged FNLM of the right half $A$ of an open hopping chain with $L=600$ and $J=1$, after a quench from a left half in $|1000\cdots\rangle$ and an empty right half, for $\gamma=0.005$, $0.01$, $0.02$, $0.05$, $0.1$, $0.5$ (from dark to light)}. Symbols: quantum-jump averages with standard errors over $N_{\rm traj}=100$ trajectories; lines: Eq.~\eqref{eq:renewal}, with the multiplet term restricted to emission points in the left half and $\rho(x,t)$ from Eq.~\eqref{eq:kinetic}, with no free parameters. Inset: unitary dynamics ($\gamma=0$), exact (symbols) and multiplet prediction (line), against $Jt$.
\label{fig:domain_wall}
\end{figure}
 
\paragraph{Consistency checks.---}
The ansatz reproduces three limits exactly. (i) Without monitoring, Eq.~\eqref{eq:hybrid} reduces to Eq.~\eqref{eq:multiplet}, which captures the exact Gaussian dynamics at leading order~\cite{SM}. (ii) If the initial correlations are themselves pairs carrying the weight $\ms(\rho)$, Eq.~\eqref{eq:hybrid} reduces to the collapsed-pair prediction of Ref.~\cite{Cao2019}, with the entanglement weight replaced by $\ms(\rho)$~\cite{SM}. (iii) The local filling entering postulate (iii) follows from an exact equation for the averaged correlations, Eq.~\eqref{eq:kinetic}, derived from the jump update~\cite{SM}.

\begin{figure*}[t!]
\centering
\includegraphics[width=\textwidth]{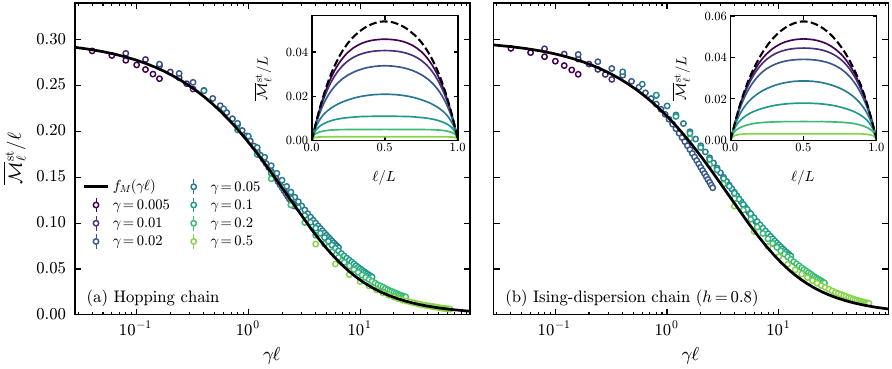}
\caption{Stationary nonlocal magic for the period-four initial state in (a) the hopping chain and (b) the Ising-dispersion chain ($h=0.8$), with $L=512$ and $J=1$. The rescaled data $\avg{\Mf}^{\rm st}_\ell/\ell$ are plotted against $\gamma\ell$ for $\gamma=0.005$, $0.01$, $0.02$, $0.05$, $0.1$, $0.2$, $0.5$ (symbols), with $\ell=8,12,\dots,128$ ($\ell\leq32$ for $\gamma=0.005$ and $0.01$, where $\ell^*=\bar v/\gamma$ is comparable to $L$ and finite-size effects appear). Data are averaged over $t\in[5/\gamma,10/\gamma]$ and $N_{\rm traj}=200$ trajectories, with error bars denoting the standard error. Solid lines show $f_M(\gamma\ell)$ from Eq.~\eqref{eq:fM}, which describes the crossover from volume-law to area-law scaling. The insets show $\avg{\Mf}^{\rm st}_\ell/L$ against $\ell/L$, with dashed lines indicating the unmonitored reference.}
\label{fig:scaling_analysis}
\end{figure*}

\paragraph{Homogeneous quench.---}
In Fig.~\ref{fig:four_models_magic} we compare Eq.~\eqref{eq:hybrid} with quantum-jump simulations. For both dispersions and over two decades of $\gamma$, the FNLM initially grows linearly in time and then saturates. Within the collapse picture, this behavior results from the competition between the ballistic spreading of quasiparticle correlations and their renewal by measurements, and the theory captures both regimes quantitatively. The only discrepancies are the microscopic oscillations visible at early times, which lie beyond the semiclassical description. For periodic initial states with $p=3,5,6,7$, the theory captures the dynamics at different monitoring rates as well~\cite{SM}; since $\ms(1/p)\neq\wt(1/p)$ for these periods, this provides a direct test of postulate (iii).

\paragraph{Inhomogeneous quenches.---}
Since the theory is local, it also applies when the initial state is periodic only locally. For a two-reservoir quench, joining two period-four halves with shifted occupation patterns, both halves have the same filling and, locally, the same rank-one multiplet structure, so that Eq.~\eqref{eq:hybrid} applies unchanged and describes the data for both dispersions~\cite{SM}. A more stringent test is a domain-wall quench, in which the left half of an open chain is periodic and the right half is empty i.e. $|10001000\cdots 1000\rangle \otimes|00000000\cdots0000\rangle$, and we compute the FNLM of the right half. Without monitoring, multiplets emitted near the junction send some of their members into the empty half, so that the FNLM grows linearly in time; Eq.~\eqref{eq:multiplet}, with the emission points restricted to the left half, reproduces this growth [Fig.~\ref{fig:domain_wall}, inset]. Under monitoring, the filling is no longer uniform, since particles spread into the empty half: $\rho(x,t)$ follows from Eq.~\eqref{eq:kinetic} with the step profile $\rho(x,0)=\frac14$ for $x<0$ and $0$ otherwise, and now depends on position and time. Eq.~~\eqref{eq:renewal} then gives the quasiparticle prediction, which captures the quantum-jump data at all monitoring rates (Fig.~\ref{fig:domain_wall}). Note that pairs renewed where the chain is still empty carry no FNLM, since $\ms(0)=0$.

\paragraph{Stationary state.---}
At long times, only the renewal term of Eq.~\eqref{eq:hybrid} survives, and we obtain
\begin{align}
\avg{\Mf}^{\rm st}_\ell&\simeq\ell\,f_M(\gamma\ell),
\label{eq:scaling}\\
f_M(x)&=\ms(\rho)\int_{-\pi}^{\pi}\frac{dk}{2\pi}\,\frac{2|v_k|}{x}\Big(1-e^{-x/2|v_k|}\Big).
\label{eq:fM}
\end{align}
Defining $\bar v=\int\frac{dk}{2\pi}2|v_k|$, which is $4J/\pi$ for the hopping chain, one finds two regimes. For $\gamma\ell\ll\bar v$, $f_M\to\ms(\rho)$, and the stationary FNLM follows a volume law, $\ell\,\ms(\rho)$, which for the period-four state coincides with the unitary saturation value $\ell\mu_4(1)$. This coincidence is specific to $p=4$: for $p\neq4$, the renewal weight $\ms(1/p)$ differs from the unitary saturation density $\wt(1/p)$ by a factor between $1.5$ and $2.6$, and the weight extracted from the quantum-jump data follows $\ms(1/p)$ to within about $5\%$~\cite{SM}. The limits $t\to\infty$ and $\gamma\to0$ therefore do not commute: arbitrarily weak monitoring changes the stationary density of nonlocal magic from $\wt(\rho)$ to $\ms(\rho)$. For $\gamma\ell\gg\bar v$, instead, $\avg{\Mf}^{\rm st}_\ell\to\ms(\rho)\,\bar v/\gamma$, i.e., an area law. The two laws meet at $\ell^*=\bar v/\gamma$~\cite{SM}. The ansatz therefore predicts that volume-law FNLM is destroyed by arbitrarily weak monitoring, through a crossover at $\ell^*$ rather than a transition, as found for the entanglement of U(1)-symmetric monitored fermions~\cite{Poboiko2023,Nie25}. As shown in Fig.~\ref{fig:scaling_analysis}, the data for $\avg{\Mf}^{\rm st}_\ell/\ell$ collapse onto $f_M(\gamma\ell)$ for both dispersions.

At half filling, the initial multiplets generate no FNLM at leading order. The N\'eel state displays the same crossover scale $\ell^*\propto1/\gamma$, but its stationary FNLM cannot be reproduced by any pair weight~\cite{SM}: there, the renewed correlations are not captured by our picture.

\paragraph{Conclusion.---}
We have shown that the dynamics of fermionic nonlocal magic under quantum-jump monitoring admits a simple, parameter-free quasiparticle-collapse description. In this picture, the quasiparticle velocities determine how correlations spread across a subsystem, while a separate weight specifies how much nonlocal magic they carry. This separation makes nonlocal magic a genuinely different probe from entanglement: the same multiplets that make the entanglement grow monotonically produce a transient dip in the nonlocal magic. The initial correlations propagate as multiplets and are renewed by quantum jumps, at a rate fixed by the jump statistics, into pairs that carry the nonlocal magic weight $\ms(\rho)$ of the local filling, which differs from the unitary weight $\wt(\rho)$, so that the limits of long times and weak monitoring do not commute. Without free parameters, this construction reproduces quantum-jump simulations of two number-conserving chains for periodic initial states at several fillings, for a two-reservoir state with shifted occupation patterns, and for a domain-wall state, where the local filling itself evolves in time; it predicts that even arbitrarily weak monitoring makes the stationary nonlocal magic cross over from a volume law to an area law at the length $\ell^*=\bar v/\gamma$. Since the construction requires only the multiplet content of the initial state and the local filling, we expect it to apply to any initial state whose multiplets generate nonlocal magic at leading order as they spread. Although the postulates of the ansatz are heuristic rather than derived, the theory contains no fitting parameters, so that its agreement with the numerics is a stringent test. At half filling, where the initial multiplets generate no nonlocal magic at leading order, the renewed correlations are not captured by our picture; this is not surprising, since even without monitoring the FNLM of the N\'eel state lies beyond the leading quasiparticle description~\cite{collura2026}. Since the FNLM depends only on two-point functions, the multiplet dip of the unmonitored dynamics could be probed with fermionic shadow tomography~\cite{ZhaoRubinMiyake2021}.

Natural directions for future investigation include a microscopic derivation of the renewal weight $\ms(\rho)$, the statistics of renewed correlations beyond the quasiparticle picture, for instance through replica field theories~\cite{Poboiko2023,Fava2023,gal2025replicafieldtheory}, and the extension of the collapse picture to jump processes that do not conserve particle number. In the transverse-field Ising chain with physical occupation monitoring, for instance, a non-Hermitian version of the unitary quasiparticle theory captures the transient no-click dynamics (Appendix~\ref{app:noclick}), whereas the dynamics with jumps lies beyond the present ansatz~\cite{Gal23,gal2025replicafieldtheory,KloiberTollinger2025}.

\paragraph{Acknowledgments.---}
The authors thank Xiangyu Cao for insightful suggestions, and Emanuele Tirrito for discussions and collaboration on related projects. This work was supported by the QUASAR-FVG project funded within the ``Piano Sviluppo e Coesione FVG'' (M.C.).

\paragraph{Data availability.---}
The data that support the findings of this article, together with the code used to generate them, are available upon request.

\bibliography{references}

@misc{SM,
  note = {See Supplemental Material for the two-mode normal form, the model details, the multiplet construction and its extension to general and inhomogeneous initial states, the derivation of the jump update and of the kinetic equation, the stationary limit and finite-ring corrections, half filling, other periods, the inhomogeneous quenches, the Ising chain with physical monitoring, and the numerical methods},
  year = {2026}
}

@misc{Gottesman1998,
  author        = {Gottesman, Daniel},
  title         = {The {H}eisenberg Representation of Quantum Computers},
  year          = {1998},
  eprint        = {quant-ph/9807006},
  archivePrefix = {arXiv},
  doi           = {10.48550/arXiv.quant-ph/9807006}
}

@article{Aaronson2004,
  author  = {Aaronson, Scott and Gottesman, Daniel},
  title   = {Improved simulation of stabilizer circuits},
  journal = {Phys. Rev. A},
  volume  = {70},
  pages   = {052328},
  year    = {2004},
  doi     = {10.1103/PhysRevA.70.052328}
}

@article{Bravyi_2005,
  author  = {Bravyi, Sergey and Kitaev, Alexei},
  title   = {Universal quantum computation with ideal {C}lifford gates and noisy ancillas},
  journal = {Phys. Rev. A},
  volume  = {71},
  pages   = {022316},
  year    = {2005},
  doi     = {10.1103/PhysRevA.71.022316}
}

@article{Veitch2014,
  author  = {Veitch, Victor and Mousavian, S. A. Hamed and Gottesman, Daniel and Emerson, Joseph},
  title   = {The resource theory of stabilizer quantum computation},
  journal = {New J. Phys.},
  volume  = {16},
  pages   = {013009},
  year    = {2014},
  doi     = {10.1088/1367-2630/16/1/013009}
}

@article{Howard2017,
  author  = {Howard, Mark and Campbell, Earl},
  title   = {Application of a Resource Theory for Magic States to Fault-Tolerant Quantum Computing},
  journal = {Phys. Rev. Lett.},
  volume  = {118},
  pages   = {090501},
  year    = {2017},
  doi     = {10.1103/PhysRevLett.118.090501}
}

@article{BravyiSmithSmolin2016,
  author  = {Bravyi, Sergey and Smith, Graeme and Smolin, John A.},
  title   = {Trading Classical and Quantum Computational Resources},
  journal = {Phys. Rev. X},
  volume  = {6},
  pages   = {021043},
  year    = {2016},
  doi     = {10.1103/PhysRevX.6.021043}
}

@article{Bravyi2019simulation,
  author  = {Bravyi, Sergey and Browne, Dan and Calpin, Padraic and Campbell, Earl and Gosset, David and Howard, Mark},
  title   = {Simulation of quantum circuits by low-rank stabilizer decompositions},
  journal = {Quantum},
  volume  = {3},
  pages   = {181},
  year    = {2019},
  doi     = {10.22331/q-2019-09-02-181}
}

@article{Beverland2020,
  author  = {Beverland, Michael and Campbell, Earl and Howard, Mark and Kliuchnikov, Vadym},
  title   = {Lower bounds on the non-{C}lifford resources for quantum computations},
  journal = {Quantum Sci. Technol.},
  volume  = {5},
  pages   = {035009},
  year    = {2020},
  doi     = {10.1088/2058-9565/ab8963}
}

@article{WangWildeSu2020,
  author  = {Wang, Xin and Wilde, Mark M. and Su, Yuan},
  title   = {Efficiently Computable Bounds for Magic State Distillation},
  journal = {Phys. Rev. Lett.},
  volume  = {124},
  pages   = {090505},
  year    = {2020},
  doi     = {10.1103/PhysRevLett.124.090505}
}

@article{Seddon2021,
  author  = {Seddon, James R. and Regula, Bartosz and Pashayan, Hakop and Ouyang, Yingkai and Campbell, Earl T.},
  title   = {Quantifying Quantum Speedups: Improved Classical Simulation From Tighter Magic Monotones},
  journal = {PRX Quantum},
  volume  = {2},
  pages   = {010345},
  year    = {2021},
  doi     = {10.1103/PRXQuantum.2.010345}
}

@article{Liu2022,
  author  = {Liu, Zi-Wen and Winter, Andreas},
  title   = {Many-Body Quantum Magic},
  journal = {PRX Quantum},
  volume  = {3},
  pages   = {020333},
  year    = {2022},
  doi     = {10.1103/PRXQuantum.3.020333}
}

@article{Leo21b,
  author        = {Leone, Lorenzo and Oliviero, Salvatore F. E. and Hamma, Alioscia},
  title         = {Stabilizer {R}{\'e}nyi Entropy},
  journal       = {Phys. Rev. Lett.},
  volume        = {128},
  pages         = {050402},
  year          = {2022},
  doi           = {10.1103/PhysRevLett.128.050402},
  eprint        = {2106.12587},
  archivePrefix = {arXiv}
}

@article{LeoneBittel2024,
  author  = {Leone, Lorenzo and Bittel, Lennart},
  title   = {Stabilizer entropies are monotones for magic-state resource theory},
  journal = {Phys. Rev. A},
  volume  = {110},
  pages   = {L040403},
  year    = {2024},
  doi     = {10.1103/PhysRevA.110.L040403}
}

@article{HaugPiroli2023,
  author  = {Haug, Tobias and Piroli, Lorenzo},
  title   = {Stabilizer entropies and nonstabilizerness monotones},
  journal = {Quantum},
  volume  = {7},
  pages   = {1092},
  year    = {2023},
  doi           = {10.22331/q-2023-08-28-1092}
}

@article{Niroula2024,
  author  = {Niroula, Pradeep and White, Christopher David and Wang, Qingfeng and Johri, Sonika and Zhu, Daiwei and Monroe, Christopher and Noel, Crystal and Gullans, Michael J.},
  title   = {Phase transition in magic with random quantum circuits},
  journal = {Nat. Phys.},
  volume  = {20},
  pages   = {1786},
  year    = {2024},
  doi           = {10.1038/s41567-024-02637-3}
}

@article{Bejan2024,
  author  = {Bejan, Mircea and McLauchlan, Campbell and B{\'e}ri, Benjamin},
  title   = {Dynamical Magic Transitions in Monitored {C}lifford+{T} Circuits},
  journal = {PRX Quantum},
  volume  = {5},
  pages   = {030332},
  year    = {2024},
  doi     = {10.1103/PRXQuantum.5.030332}
}

@article{Fux2024,
  author  = {Fux, Gerald E. and Tirrito, Emanuele and Dalmonte, Marcello and Fazio, Rosario},
  title   = {Entanglement--nonstabilizerness separation in hybrid quantum circuits},
  journal = {Phys. Rev. Res.},
  volume  = {6},
  pages   = {L042030},
  year    = {2024},
  doi     = {10.1103/PhysRevResearch.6.L042030}
}

@article{TarabungaTirrito2024,
  author        = {Tarabunga, Poetri Sonya and Tirrito, Emanuele},
  title         = {Magic transition in measurement-only circuits},
  journal       = {npj Quantum Inf.},
  volume        = {11},
  pages         = {166},
  year          = {2025},
  eprint        = {2407.15939},
  archivePrefix = {arXiv},
  doi           = {10.1038/s41534-025-01104-y}
}

@article{cao2024,
  author        = {Cao, ChunJun},
  title         = {Non-trivial area operators require non-local magic},
  journal       = {J. High Energy Phys.},
  volume        = {2024},
  number        = {11},
  pages         = {105},
  year          = {2024},
  doi           = {10.1007/JHEP11(2024)105},
  eprint        = {2306.14996},
  archivePrefix = {arXiv}
}

@article{cao25,
  author  = {Cao, ChunJun and Cheng, Gong and Hamma, Alioscia and Leone, Lorenzo and Munizzi, William and Oliviero, Salvatore F. E.},
  title   = {Gravitational Backreaction is Magical},
  journal = {PRX Quantum},
  volume  = {6},
  pages   = {040375},
  year    = {2025},
  doi     = {10.1103/z3vr-w5c5}
}

@article{Qian_2025,
  author        = {Qian, Dongheng and Wang, Jing},
  title         = {Quantum nonlocal nonstabilizerness},
  journal       = {Phys. Rev. A},
  volume        = {111},
  pages         = {052443},
  year          = {2025},
  doi           = {10.1103/PhysRevA.111.052443},
  eprint        = {2502.06393},
  archivePrefix = {arXiv}
}

@article{Andreadakis2026,
  author  = {Andreadakis, Faidon and Zanardi, Paolo},
  title   = {Exact link between non-local nonstabilizerness and operator entanglement},
  journal = {Phys. Rev. A},
  volume  = {113},
  pages   = {L010404},
  year    = {2026},
  doi     = {10.1103/9x56-2b45}
}

@article{Cepollaro2025,
  author  = {Cepollaro, Simone and Cusumano, Stefano and Hamma, Alioscia and Lo Giudice, Giuseppe and Odavi{\'c}, Jovan},
  title   = {Harvesting stabilizer entropy and nonlocality from a quantum field},
  journal = {Phys. Rev. D},
  volume  = {112},
  pages   = {105012},
  year    = {2025},
  doi     = {10.1103/4brj-cl26}
}

@article{Robin2026,
  author  = {Robin, Caroline E. P. and Savage, Martin J.},
  title   = {Antiflatness and nonlocal nonstabilizerness in two-particle scattering processes},
  journal = {Phys. Rev. D},
  volume  = {114},
  pages   = {014007},
  year    = {2026},
  doi     = {10.1103/hvft-hk9p}
}

@misc{Ahmad2025,
  author        = {Ahmad, H. G. and Esposito, G. and Stasino, V. and Odavi{\'c}, J. and Cosenza, C. and Sarno, A. and Mastrovito, P. and Viscardi, M. and Cusumano, S. and Tafuri, F. and Massarotti, D. and Hamma, A.},
  title         = {Experimental demonstration of non-local magic in a superconducting quantum processor},
  year          = {2025},
  eprint        = {2511.15576},
  archivePrefix = {arXiv},
  primaryClass  = {quant-ph},
  doi           = {10.48550/arXiv.2511.15576}
}

@misc{sierant2026,
  author        = {Sierant, Piotr},
  title         = {Spectral geometry of nonlocal stabilizer entropy},
  year          = {2026},
  eprint        = {2609.13106},
  archivePrefix = {arXiv},
  primaryClass  = {quant-ph},
  doi           = {10.48550/arXiv.2609.13106}
}

@misc{Sierant2026exact,
  author        = {Sierant, Piotr},
  title         = {Exact quantification of nonlocal magic},
  year          = {2026},
  eprint        = {2608.28563},
  archivePrefix = {arXiv},
  primaryClass  = {quant-ph},
  doi           = {10.48550/arXiv.2608.28563}
}

@misc{Huang2026,
  author        = {Huang, Xuyang and Chen, Gang and Yao, Yao},
  title         = {Intrinsic spectral structure of bipartite nonlocal magic resource},
  year          = {2026},
  eprint        = {2606.24368},
  archivePrefix = {arXiv},
  primaryClass  = {quant-ph},
  doi           = {10.48550/arXiv.2606.24368}
}

@misc{Torre2026,
  author        = {Torre, Gianpaolo and Franchini, Fabio and Giampaolo, Salvatore Marco},
  title         = {Nonlocal magic from the entanglement spectrum},
  year          = {2026},
  eprint        = {2607.07808},
  archivePrefix = {arXiv},
  primaryClass  = {quant-ph},
  doi           = {10.48550/arXiv.2607.07808}
}

@misc{Liu2026,
  author        = {Liu, L.-Y.-N. and Cui, Jian},
  title         = {Nonlocal nonstabilizerness for slightly entangled quantum many-body states},
  year          = {2026},
  eprint        = {2607.10714},
  archivePrefix = {arXiv},
  primaryClass  = {quant-ph},
  doi           = {10.48550/arXiv.2607.10714}
}

@misc{Franchini2026,
  author        = {Franchini, Fabio and Giampaolo, Salvatore Marco},
  title         = {Schmidt-gauge non-local magic: Representation, optimality, and mathematical properties},
  year          = {2026},
  eprint        = {2608.02745},
  archivePrefix = {arXiv},
  primaryClass  = {quant-ph},
  doi           = {10.48550/arXiv.2608.02745}
}

@misc{Aditya2026,
  author        = {Aditya, Sreemayee and Sierant, Piotr and Turkeshi, Xhek},
  title         = {Nonlocal Magic Spreading in Many-body Quantum Dynamics: From Chaotic Evolution to Quasi-particle Picture in Integrable Models},
  year          = {2026},
  eprint        = {2609.20951},
  archivePrefix = {arXiv},
  primaryClass  = {quant-ph},
  doi           = {10.48550/arXiv.2609.20951}
}

@article{Valiant2002,
  author  = {Valiant, Leslie G.},
  title   = {Quantum Circuits That Can Be Simulated Classically in Polynomial Time},
  journal = {SIAM J. Comput.},
  volume  = {31},
  pages   = {1229},
  year    = {2002},
  doi     = {10.1137/S0097539700377025}
}

@article{TerhalDiVincenzo2002,
  author  = {Terhal, Barbara M. and DiVincenzo, David P.},
  title   = {Classical simulation of noninteracting-fermion quantum circuits},
  journal = {Phys. Rev. A},
  volume  = {65},
  pages   = {032325},
  year    = {2002},
  doi     = {10.1103/PhysRevA.65.032325}
}

@misc{Knill2001,
  author        = {Knill, Emanuel},
  title         = {Fermionic Linear Optics and Matchgates},
  year          = {2001},
  eprint        = {quant-ph/0108033},
  archivePrefix = {arXiv},
  doi           = {10.48550/arXiv.quant-ph/0108033}
}

@article{JozsaMiyake2008,
  author  = {Jozsa, Richard and Miyake, Akimasa},
  title   = {Matchgates and classical simulation of quantum circuits},
  journal = {Proc. R. Soc. A},
  volume  = {464},
  pages   = {3089},
  year    = {2008},
  doi     = {10.1098/rspa.2008.0189}
}

@article{Bravyi2005,
  author        = {Bravyi, Sergey},
  title         = {Lagrangian representation for fermionic linear optics},
  journal       = {Quantum Inf. Comput.},
  volume        = {5},
  pages         = {216},
  year          = {2005},
  eprint        = {quant-ph/0404180},
  archivePrefix = {arXiv},
  doi           = {10.48550/arXiv.quant-ph/0404180}
}

@article{Hebenstreit2019,
  author  = {Hebenstreit, M. and Jozsa, R. and Kraus, B. and Strelchuk, S. and Yoganathan, M.},
  title   = {All Pure Fermionic Non-{G}aussian States Are Magic States for Matchgate Computations},
  journal = {Phys. Rev. Lett.},
  volume  = {123},
  pages   = {080503},
  year    = {2019},
  doi     = {10.1103/PhysRevLett.123.080503}
}

@article{Oszmaniec2022,
  author  = {Oszmaniec, Micha{\l} and Dangniam, Ninnat and Morales, Mauro E. S. and Zimbor{\'a}s, Zolt{\'a}n},
  title   = {Fermion Sampling: A Robust Quantum Computational Advantage Scheme Using Fermionic Linear Optics and Magic Input States},
  journal = {PRX Quantum},
  volume  = {3},
  pages   = {020328},
  year    = {2022},
  doi     = {10.1103/PRXQuantum.3.020328}
}

@article{DiasKoenig2024,
  author  = {Dias, Beatriz and Koenig, Robert},
  title   = {Classical simulation of non-{G}aussian fermionic circuits},
  journal = {Quantum},
  volume  = {8},
  pages   = {1350},
  year    = {2024},
  doi           = {10.22331/q-2024-05-21-1350}
}

@article{Sierant2026FMR,
  author  = {Sierant, Piotr and Stornati, Paolo and Turkeshi, Xhek},
  title   = {Fermionic Magic Resources of Quantum Many-Body Systems},
  journal = {PRX Quantum},
  volume  = {7},
  pages   = {010302},
  year    = {2026},
  doi     = {10.1103/3yx4-1j27}
}

@article{Collura2026quantum,
  author        = {Collura, Mario and De Nardis, Jacopo and Alba, Vincenzo and Lami, Guglielmo},
  title         = {The non-stabilizerness of fermionic {G}aussian states},
  journal       = {Quantum},
  volume        = {10},
  pages         = {2036},
  year          = {2026},
  doi           = {10.22331/q-2026-03-23-2036},
  eprint        = {2412.05367},
  archivePrefix = {arXiv}
}

@misc{Iannotti2026,
  author        = {Iannotti, Daniele and Magni, Beatrice and Cioli, Riccardo and Hamma, Alioscia and Turkeshi, Xhek},
  title         = {Non-Local Magic Resources for Fermionic {G}aussian States},
  year          = {2026},
  eprint        = {2604.27049},
  archivePrefix = {arXiv},
  primaryClass  = {quant-ph},
  doi           = {10.48550/arXiv.2604.27049}
}

@misc{collura2026,
  author        = {Collura, Mario and B{\'e}ri, Benjamin and Tirrito, Emanuele},
  title         = {Nonlocal nonstabilizerness in free fermion models},
  year          = {2026},
  eprint        = {2604.27055},
  archivePrefix = {arXiv},
  primaryClass  = {quant-ph},
  doi           = {10.48550/arXiv.2604.27055}
}

@article{ZhaoRubinMiyake2021,
  author  = {Zhao, Andrew and Rubin, Nicholas C. and Miyake, Akimasa},
  title   = {Fermionic Partial Tomography via Classical Shadows},
  journal = {Phys. Rev. Lett.},
  volume  = {127},
  pages   = {110504},
  year    = {2021},
  doi     = {10.1103/PhysRevLett.127.110504}
}

@article{SuraceTagliacozzo2022,
  author  = {Surace, Jacopo and Tagliacozzo, Luca},
  title   = {Fermionic {G}aussian states: an introduction to numerical approaches},
  journal = {SciPost Phys. Lect. Notes},
  pages   = {54},
  year    = {2022},
  doi     = {10.21468/SciPostPhysLectNotes.54}
}

@article{Peschel2003,
  author  = {Peschel, Ingo},
  title   = {Calculation of reduced density matrices from correlation functions},
  journal = {J. Phys. A: Math. Gen.},
  volume  = {36},
  pages   = {L205},
  year    = {2003},
  doi     = {10.1088/0305-4470/36/14/101}
}

@article{Vidal2003,
  author  = {Vidal, G. and Latorre, J. I. and Rico, E. and Kitaev, A.},
  title   = {Entanglement in Quantum Critical Phenomena},
  journal = {Phys. Rev. Lett.},
  volume  = {90},
  pages   = {227902},
  year    = {2003},
  doi     = {10.1103/PhysRevLett.90.227902}
}

@article{PeschelEisler2009,
  author  = {Peschel, Ingo and Eisler, Viktor},
  title   = {Reduced density matrices and entanglement entropy in free lattice models},
  journal = {J. Phys. A: Math. Theor.},
  volume  = {42},
  pages   = {504003},
  year    = {2009},
  doi     = {10.1088/1751-8113/42/50/504003}
}

@article{SantoroIsing,
  author  = {Mbeng, Glen Bigan and Russomanno, Angelo and Santoro, Giuseppe E.},
  title   = {The quantum {I}sing chain for beginners},
  journal = {SciPost Phys. Lect. Notes},
  pages   = {82},
  year    = {2024},
  doi     = {10.21468/SciPostPhysLectNotes.82}
}

@article{CalabreseCardy2005,
  author  = {Calabrese, Pasquale and Cardy, John},
  title   = {Evolution of entanglement entropy in one-dimensional systems},
  journal = {J. Stat. Mech.},
  volume  = {2005},
  pages   = {P04010},
  year    = {2005},
  doi     = {10.1088/1742-5468/2005/04/P04010}
}

@article{Fag2008,
  author  = {Fagotti, Maurizio and Calabrese, Pasquale},
  title   = {Evolution of entanglement entropy following a quantum quench: Analytic results for the {XY} chain in a transverse magnetic field},
  journal = {Phys. Rev. A},
  volume  = {78},
  pages   = {010306(R)},
  year    = {2008},
  doi     = {10.1103/PhysRevA.78.010306}
}

@article{AlbaCalabrese2017,
  author  = {Alba, Vincenzo and Calabrese, Pasquale},
  title   = {Entanglement and thermodynamics after a quantum quench in integrable systems},
  journal = {Proc. Natl. Acad. Sci. U.S.A.},
  volume  = {114},
  pages   = {7947},
  year    = {2017},
  doi     = {10.1073/pnas.1703516114}
}

@article{AlbaCalabrese2018,
  author  = {Alba, Vincenzo and Calabrese, Pasquale},
  title   = {Entanglement dynamics after quantum quenches in generic integrable systems},
  journal = {SciPost Phys.},
  volume  = {4},
  pages   = {017},
  year    = {2018},
  doi     = {10.21468/SciPostPhys.4.3.017}
}

@article{Calabrese2020notes,
  author  = {Calabrese, Pasquale},
  title   = {Entanglement spreading in non-equilibrium integrable systems},
  journal = {SciPost Phys. Lect. Notes},
  pages   = {20},
  year    = {2020},
  doi     = {10.21468/SciPostPhysLectNotes.20}
}

@article{Bertini2018,
  author  = {Bertini, Bruno and Tartaglia, Elena and Calabrese, Pasquale},
  title   = {Entanglement and diagonal entropies after a quench with no pair structure},
  journal = {J. Stat. Mech.},
  volume  = {2018},
  pages   = {063104},
  year    = {2018},
  doi     = {10.1088/1742-5468/aac73f}
}

@article{Bertini_2018,
  author  = {Bertini, Bruno and Fagotti, Maurizio and Piroli, Lorenzo and Calabrese, Pasquale},
  title   = {Entanglement evolution and generalised hydrodynamics: noninteracting systems},
  journal = {J. Phys. A: Math. Theor.},
  volume  = {51},
  pages   = {39LT01},
  year    = {2018},
  doi     = {10.1088/1751-8121/aad82e}
}

@article{Alba2018,
  author  = {Alba, Vincenzo},
  title   = {Entanglement and quantum transport in integrable systems},
  journal = {Phys. Rev. B},
  volume  = {97},
  pages   = {245135},
  year    = {2018},
  doi     = {10.1103/PhysRevB.97.245135}
}

@article{Bastianello_2018,
  author  = {Bastianello, Alvise and Calabrese, Pasquale},
  title   = {Spreading of entanglement and correlations after a quench with intertwined quasiparticles},
  journal = {SciPost Phys.},
  volume  = {5},
  pages   = {033},
  year    = {2018},
  doi     = {10.21468/SciPostPhys.5.4.033}
}

@article{BastianelloCollura2020,
  author        = {Bastianello, Alvise and Collura, Mario},
  title         = {Entanglement spreading and quasiparticle picture beyond the pair structure},
  journal       = {SciPost Phys.},
  volume        = {8},
  pages         = {045},
  year          = {2020},
  doi           = {10.21468/SciPostPhys.8.3.045},
  eprint        = {2001.01671},
  archivePrefix = {arXiv}
}

@article{Fagotti2017,
  author  = {Fagotti, Maurizio},
  title   = {Higher-order generalized hydrodynamics in one dimension: The noninteracting test},
  journal = {Phys. Rev. B},
  volume  = {96},
  pages   = {220302},
  year    = {2017},
  doi     = {10.1103/PhysRevB.96.220302}
}

@article{Car21,
  author        = {Carollo, Federico and Alba, Vincenzo},
  title         = {Dissipative Quasiparticle Picture for Quadratic {M}arkovian Open Quantum Systems},
  journal       = {Phys. Rev. B},
  volume        = {105},
  pages         = {144305},
  year          = {2022},
  doi           = {10.1103/PhysRevB.105.144305},
  eprint        = {2106.11997},
  archivePrefix = {arXiv}
}

@article{AlbaCarollo2021,
  author  = {Alba, Vincenzo and Carollo, Federico},
  title   = {Spreading of correlations in {M}arkovian open quantum systems},
  journal = {Phys. Rev. B},
  volume  = {103},
  pages   = {L020302},
  year    = {2021},
  doi     = {10.1103/PhysRevB.103.L020302}
}

@article{Car22,
  author  = {Carollo, Federico and Alba, Vincenzo},
  title   = {Entangled Multiplets and Spreading of Quantum Correlations in a Continuously Monitored Tight-Binding Chain},
  journal = {Phys. Rev. B},
  volume  = {106},
  pages   = {L220304},
  year    = {2022},
  doi     = {10.1103/PhysRevB.106.L220304}
}

@article{Eisler2011,
  author  = {Eisler, Viktor},
  title   = {Crossover between ballistic and diffusive transport: the quantum exclusion process},
  journal = {J. Stat. Mech.},
  volume  = {2011},
  pages   = {P06007},
  year    = {2011},
  doi           = {10.1088/1742-5468/2011/06/P06007}
}

@article{Znidaric2010,
  author  = {{\v{Z}}nidari{\v{c}}, Marko},
  title   = {Exact solution for a diffusive nonequilibrium steady state of an open quantum chain},
  journal = {J. Stat. Mech.},
  volume  = {2010},
  pages   = {L05002},
  year    = {2010},
  doi           = {10.1088/1742-5468/2010/05/L05002}
}

@article{BernardJin2019,
  author  = {Bernard, Denis and Jin, Tony},
  title   = {Open Quantum Symmetric Simple Exclusion Process},
  journal = {Phys. Rev. Lett.},
  volume  = {123},
  pages   = {080601},
  year    = {2019},
  doi     = {10.1103/PhysRevLett.123.080601}
}

@article{DalibardCastinMolmer1992,
  author  = {Dalibard, Jean and Castin, Yvan and M{\o}lmer, Klaus},
  title   = {Wave-function approach to dissipative processes in quantum optics},
  journal = {Phys. Rev. Lett.},
  volume  = {68},
  pages   = {580},
  year    = {1992},
  doi     = {10.1103/PhysRevLett.68.580}
}

@article{Gardiner1992,
  author  = {Gardiner, C. W. and Parkins, A. S. and Zoller, P.},
  title   = {Wave-function quantum stochastic differential equations and quantum-jump simulation methods},
  journal = {Phys. Rev. A},
  volume  = {46},
  pages   = {4363},
  year    = {1992},
  doi     = {10.1103/PhysRevA.46.4363}
}

@article{PlenioKnight1998,
  author  = {Plenio, M. B. and Knight, P. L.},
  title   = {The quantum-jump approach to dissipative dynamics in quantum optics},
  journal = {Rev. Mod. Phys.},
  volume  = {70},
  pages   = {101},
  year    = {1998},
  doi     = {10.1103/RevModPhys.70.101}
}

@article{Daley2014,
  author  = {Daley, Andrew J.},
  title   = {Quantum trajectories and open many-body quantum systems},
  journal = {Adv. Phys.},
  volume  = {63},
  pages   = {77},
  year    = {2014},
  doi     = {10.1080/00018732.2014.933502}
}

@book{WisemanMilburn2009,
  author    = {Wiseman, Howard M. and Milburn, Gerard J.},
  title     = {Quantum Measurement and Control},
  publisher = {Cambridge University Press},
  address   = {Cambridge},
  year      = {2009},
  doi           = {10.1017/CBO9780511813948}
}

@article{Skinner2019,
  author  = {Skinner, Brian and Ruhman, Jonathan and Nahum, Adam},
  title   = {Measurement-Induced Phase Transitions in the Dynamics of Entanglement},
  journal = {Phys. Rev. X},
  volume  = {9},
  pages   = {031009},
  year    = {2019},
  doi     = {10.1103/PhysRevX.9.031009}
}

@article{Li2018,
  author  = {Li, Yaodong and Chen, Xiao and Fisher, Matthew P. A.},
  title   = {Quantum {Z}eno effect and the many-body entanglement transition},
  journal = {Phys. Rev. B},
  volume  = {98},
  pages   = {205136},
  year    = {2018},
  doi     = {10.1103/PhysRevB.98.205136}
}

@article{Li2019,
  author  = {Li, Yaodong and Chen, Xiao and Fisher, Matthew P. A.},
  title   = {Measurement-driven entanglement transition in hybrid quantum circuits},
  journal = {Phys. Rev. B},
  volume  = {100},
  pages   = {134306},
  year    = {2019},
  doi     = {10.1103/PhysRevB.100.134306}
}

@article{Chan2019,
  author  = {Chan, Amos and Nandkishore, Rahul M. and Pretko, Michael and Smith, Graeme},
  title   = {Unitary-projective entanglement dynamics},
  journal = {Phys. Rev. B},
  volume  = {99},
  pages   = {224307},
  year    = {2019},
  doi     = {10.1103/PhysRevB.99.224307}
}

@article{Fisher2023,
  author  = {Fisher, Matthew P. A. and Khemani, Vedika and Nahum, Adam and Vijay, Sagar},
  title   = {Random Quantum Circuits},
  journal = {Annu. Rev. Condens. Matter Phys.},
  volume  = {14},
  pages   = {335},
  year    = {2023},
  doi     = {10.1146/annurev-conmatphys-031720-030658}
}

@incollection{PotterVasseur2022,
  author        = {Potter, Andrew C. and Vasseur, Romain},
  title         = {Entanglement dynamics in hybrid quantum circuits},
  booktitle     = {Entanglement in Spin Chains},
  editor        = {Bayat, Abolfazl and Bose, Sougato and Johannesson, Henrik},
  publisher     = {Springer},
  address       = {Cham},
  pages         = {211},
  year          = {2022},
  eprint        = {2111.08018},
  archivePrefix = {arXiv},
  doi           = {10.1007/978-3-031-03998-0_9}
}

@article{Cao2019,
  author        = {Cao, Xiangyu and Tilloy, Antoine and De Luca, Andrea},
  title         = {Entanglement in a fermion chain under continuous monitoring},
  journal       = {SciPost Phys.},
  volume        = {7},
  pages         = {024},
  year          = {2019},
  doi           = {10.21468/SciPostPhys.7.2.024},
  eprint        = {1804.04638},
  archivePrefix = {arXiv}
}

@article{Alb20c,
  author  = {Alberton, O. and Buchhold, M. and Diehl, S.},
  title   = {Entanglement Transition in a Monitored Free-Fermion Chain: From Extended Criticality to Area Law},
  journal = {Phys. Rev. Lett.},
  volume  = {126},
  pages   = {170602},
  year    = {2021},
  doi     = {10.1103/PhysRevLett.126.170602}
}

@article{Fuj20,
  author  = {Fuji, Yohei and Ashida, Yuto},
  title   = {Measurement-Induced Quantum Criticality under Continuous Monitoring},
  journal = {Phys. Rev. B},
  volume  = {102},
  pages   = {054302},
  year    = {2020},
  doi     = {10.1103/PhysRevB.102.054302}
}

@article{Buchhold2021,
  author  = {Buchhold, M. and Minoguchi, Y. and Altland, A. and Diehl, S.},
  title   = {Effective Theory for the Measurement-Induced Phase Transition of {D}irac Fermions},
  journal = {Phys. Rev. X},
  volume  = {11},
  pages   = {041004},
  year    = {2021},
  doi     = {10.1103/PhysRevX.11.041004}
}

@article{Coppola2022,
  author  = {Coppola, M. and Tirrito, E. and Karevski, D. and Collura, M.},
  title   = {Growth of entanglement entropy under local projective measurements},
  journal = {Phys. Rev. B},
  volume  = {105},
  pages   = {094303},
  year    = {2022},
  doi     = {10.1103/PhysRevB.105.094303}
}

@article{Ladewig2022,
  author  = {Ladewig, B. and Diehl, S. and Buchhold, M.},
  title   = {Monitored open fermion dynamics: Exploring the interplay of measurement, decoherence, and free {H}amiltonian evolution},
  journal = {Phys. Rev. Res.},
  volume  = {4},
  pages   = {033001},
  year    = {2022},
  doi     = {10.1103/PhysRevResearch.4.033001}
}

@article{Poboiko2023,
  author  = {Poboiko, Igor and P{\"o}pperl, Paul and Gornyi, Igor V. and Mirlin, Alexander D.},
  title   = {Theory of Free Fermions under Random Projective Measurements},
  journal = {Phys. Rev. X},
  volume  = {13},
  pages   = {041046},
  year    = {2023},
  doi     = {10.1103/PhysRevX.13.041046}
}

@article{Nie25,
  author        = {Niederegger, Clemens and Vovk, Tatiana and Starchl, Elias and Sieberer, Lukas M.},
  title         = {Absence of measurement- and unraveling-induced entanglement transitions in continuously monitored one-dimensional free fermions},
  journal       = {Phys. Rev. B},
  volume        = {113},
  pages         = {144317},
  year          = {2026},
  doi           = {10.1103/636v-6s73},
  eprint        = {2510.19459},
  archivePrefix = {arXiv}
}

@article{Fava2023,
  author  = {Fava, Michele and Piroli, Lorenzo and Swann, Tobias and Bernard, Denis and Nahum, Adam},
  title   = {Nonlinear Sigma Models for Monitored Dynamics of Free Fermions},
  journal = {Phys. Rev. X},
  volume  = {13},
  pages   = {041045},
  year    = {2023},
  doi     = {10.1103/PhysRevX.13.041045}
}

@article{KloiberTollinger2025,
  author        = {Kloiber-Tollinger, Felix and Sieberer, Lukas M.},
  title         = {Replica {K}eldysh field theory of quantum-jump processes: General formalism and application to imbalanced and inefficient fermion counting},
  journal       = {Phys. Rev. B},
  volume        = {114},
  pages         = {024307},
  year          = {2026},
  doi           = {10.1103/pgyt-cs59},
  eprint        = {2512.16520},
  archivePrefix = {arXiv}
}

@article{Pic24,
  author        = {Piccitto, Giulia and Rossini, Davide and Russomanno, Angelo},
  title         = {The Impact of Different Unravelings in a Monitored System of Free Fermions},
  journal       = {Eur. Phys. J. B},
  volume        = {97},
  pages         = {90},
  year          = {2024},
  doi           = {10.1140/epjb/s10051-024-00725-0},
  eprint        = {2402.06597},
  archivePrefix = {arXiv}
}

@misc{Young2024,
  author        = {Young, Thomas and Gangardt, Dimitri M. and von Keyserlingk, Curt},
  title         = {Diffusive entanglement growth in a monitored harmonic chain},
  year          = {2024},
  eprint        = {2403.04022},
  archivePrefix = {arXiv},
  primaryClass  = {quant-ph},
  doi           = {10.48550/arXiv.2403.04022}
}

@misc{Delmonte2026,
  author        = {Delmonte, Anna and Schir{\`o}, Marco},
  title         = {Entanglement Dynamics across a Monitored Quantum Point Contact},
  year          = {2026},
  eprint        = {2605.22555},
  archivePrefix = {arXiv},
  primaryClass  = {cond-mat.mes-hall},
  doi           = {10.48550/arXiv.2605.22555}
}

@article{Tur21c,
  author        = {Turkeshi, Xhek and Biella, Alberto and Fazio, Rosario and Dalmonte, Marcello and Schir{\`o}, Marco},
  title         = {Measurement-Induced Entanglement Transitions in the Quantum {I}sing Chain: From Infinite to Zero Clicks},
  journal       = {Phys. Rev. B},
  volume        = {103},
  pages         = {224210},
  year          = {2021},
  doi           = {10.1103/PhysRevB.103.224210},
  eprint        = {2103.09138},
  archivePrefix = {arXiv}
}

@article{TurkeshiResetting2022,
  author        = {Turkeshi, Xhek and Dalmonte, Marcello and Fazio, Rosario and Schir{\`o}, Marco},
  title         = {Entanglement Transitions from Stochastic Resetting of Non-{H}ermitian Quasiparticles},
  journal       = {Phys. Rev. B},
  volume        = {105},
  pages         = {L241114},
  year          = {2022},
  doi           = {10.1103/PhysRevB.105.L241114},
  eprint        = {2111.03500},
  archivePrefix = {arXiv}
}

@article{TurkeshiResettingErratum2023,
  author  = {Turkeshi, Xhek and Dalmonte, Marcello and Fazio, Rosario and Schir{\`o}, Marco},
  title   = {Erratum: Entanglement Transitions from Stochastic Resetting of Non-{H}ermitian Quasiparticles},
  journal = {Phys. Rev. B},
  volume  = {107},
  pages   = {079901},
  year    = {2023},
  doi     = {10.1103/PhysRevB.107.079901}
}

@article{Tur22,
  author  = {Turkeshi, Xhek and Schir{\`o}, Marco},
  title   = {Entanglement and Correlation Spreading in Non-{H}ermitian Spin Chains},
  journal = {Phys. Rev. B},
  volume  = {107},
  pages   = {L020403},
  year    = {2023},
  doi     = {10.1103/PhysRevB.107.L020403}
}

@article{Gal22,
  author        = {Le Gal, Youenn and Turkeshi, Xhek and Schir{\`o}, Marco},
  title         = {Volume-to-Area Law Entanglement Transition in a Non-{H}ermitian Free Fermionic Chain},
  journal       = {SciPost Phys.},
  volume        = {14},
  pages         = {138},
  year          = {2023},
  doi           = {10.21468/SciPostPhys.14.5.138},
  eprint        = {2210.11937},
  archivePrefix = {arXiv}
}

@article{Gal23,
  author  = {Le Gal, Youenn and Turkeshi, Xhek and Schir{\`o}, Marco},
  title   = {Entanglement Dynamics in Monitored Systems and the Role of Quantum Jumps},
  journal = {PRX Quantum},
  volume  = {5},
  pages   = {030329},
  year    = {2024},
  doi     = {10.1103/PRXQuantum.5.030329}
}

@article{Zer23,
  author        = {Zerba, Caterina and Silva, Alessandro},
  title         = {Measurement Phase Transitions in the No-Click Limit as Quantum Phase Transitions of a Non-{H}ermitian Vacuum},
  journal       = {SciPost Phys. Core},
  volume        = {6},
  pages         = {051},
  year          = {2023},
  doi           = {10.21468/SciPostPhysCore.6.3.051},
  eprint        = {2301.07383},
  archivePrefix = {arXiv}
}

@article{BiellaSchiro2021,
  author  = {Biella, Alberto and Schir{\`o}, Marco},
  title   = {Many-Body Quantum {Z}eno Effect and Measurement-Induced Subradiance Transition},
  journal = {Quantum},
  volume  = {5},
  pages   = {528},
  year    = {2021},
  doi           = {10.22331/q-2021-08-19-528}
}

@misc{gal2025replicafieldtheory,
  author        = {Le Gal, Youenn and Schir{\`o}, Marco},
  title         = {Replica Field Theory of Quantum Jumps Monitoring: Application to the {I}sing Chain},
  year          = {2025},
  eprint        = {2511.22506},
  archivePrefix = {arXiv},
  primaryClass  = {quant-ph},
  doi           = {10.48550/arXiv.2511.22506}
}

@misc{Tir25,
  author        = {Tirrito, Emanuele and Lumia, Luca and Paviglianiti, Alessio and Lami, Guglielmo and Silva, Alessandro and Turkeshi, Xhek and Collura, Mario},
  title         = {Magic Phase Transitions in Monitored {G}aussian Fermions},
  year          = {2025},
  eprint        = {2507.07179},
  archivePrefix = {arXiv},
  primaryClass  = {quant-ph},
  doi           = {10.48550/arXiv.2507.07179}
}

@misc{wang2025,
  author        = {Wang, Cheng and Yang, Zhi-Cheng and Zhou, Tianci and Chen, Xiao},
  title         = {Magic transition in monitored free fermion dynamics},
  year          = {2025},
  eprint        = {2507.10688},
  archivePrefix = {arXiv},
  primaryClass  = {quant-ph},
  doi           = {10.48550/arXiv.2507.10688}
}

\section*{End Matter}
\appendix
\setcounter{secnumdepth}{1}
\section{No-click dynamics of the monitored Ising chain}
\label{app:noclick}

\begin{figure}[htbp]
\centering
\includegraphics[width=\linewidth]{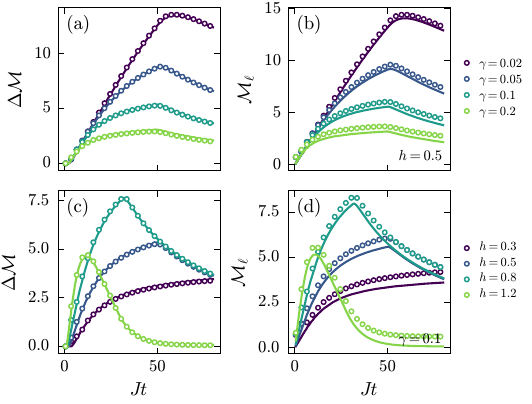}
\caption{No-click dynamics of nonlocal magic in the monitored Ising chain. The left column (a,c) shows the subtracted nonlocal magic $\Delta\Mf(t)=\Mf_\ell(t)-\Mf_{\ell_2}(t)$, and the right column (b,d) the full $\Mf_\ell(t)$. Panels (a,b) vary $\gamma$ at fixed $h=0.5$, while panels (c,d) vary $h$ at fixed $\gamma=0.1$. Solid lines represent Eq.~\eqref{eq:tfim_nc}, and open symbols the exact no-click dynamics of the infinite chain. Here $\ell=100$, $\ell_2=4$, and $J=1$.}
\label{fig:no_click_qp}
\end{figure}
We consider the transverse-field Ising chain with physical occupation monitoring, $L_j=\sqrt\gamma\,c_j^\dagger c_j$, where $c_j$ are Jordan--Wigner fermions. Since jumps now change the fermion number, the collapse ansatz does not apply in its present form~\cite{Gal23}; we therefore consider the evolution conditioned on the absence of jumps, generated by $H_{\rm eff}=H-\frac{i\gamma}{2}\sum_jc_j^\dagger c_j$~\cite{Tur21c,Tur22,Zer23,Gal22,BiellaSchiro2021}. In each sector $(k,-k)$, this yields the complex quasiparticle energy~\cite{SM}
\begin{equation}
\Omega_k=\sqrt{\epsilon_{\rm I}(k)^2-\tfrac{\gamma^2}{4}-2i\gamma(h-J\cos k)},
\label{eq:tfim_complex}
\end{equation}
where $\epsilon_{\rm I}(k)=2\sqrt{(h-J\cos k)^2+J^2\sin^2k}$ is the Ising dispersion. Since $H_{\rm eff}$ couples only the modes $k$ and $-k$, the normalized no-click state keeps the pair form $\prod_{k>0}(\alpha_k+\beta_kc_k^\dagger c_{-k}^\dagger)|0\rangle$, but the occupation $\vartheta_k^{\rm nc}(t)$ of the Ising Bogoliubov modes is no longer conserved~\cite{SM}. The quasiparticle prediction then reads
\begin{equation}
\Mf^{\rm nc}_\ell(t)\simeq\int_{-\pi}^{\pi}\frac{dk}{2\pi}\,\min\big(2|v_k^{\rm nc}|t,\ell\big)\,\wt\big[\vartheta_k^{\rm nc}(t)\big],
\label{eq:tfim_nc}
\end{equation}
with the effective group velocity
\begin{equation}
v_k^{\rm nc}=\operatorname{Re}\frac{\partial\Omega_k}{\partial k}=\operatorname{Re}\frac{J\sin k\,(4h-i\gamma)}{\Omega_k},
\label{eq:tfim_velocity}
\end{equation}
which reduces to the Ising group velocity at $\gamma=0$. The no-click trajectory occurs with a probability that decays exponentially in time and system size, and Eq.~\eqref{eq:tfim_nc} involves no collapse: it follows from the exact pair structure of the no-click state. Starting from the fermionic vacuum, i.e., the fully polarized state $|\uparrow\uparrow\cdots\rangle$, Fig.~\ref{fig:no_click_qp} compares Eq.~\eqref{eq:tfim_nc} with the exact no-click dynamics. To isolate the propagating contribution, we subtract the FNLM of a short interval, $\Delta\Mf(t)=\Mf_\ell(t)-\Mf_{\ell_2}(t)$, which removes the subleading contribution common to both sizes. The theory captures the transient growth over a broad range of $\gamma$ and $h$. At long times, instead, the no-click state approaches the vacuum of $H_{\rm eff}$, whose FNLM lies outside the quasiparticle description; the stationary state and the dynamics with jumps are discussed in Ref.~\cite{SM}.







\onecolumngrid

\begin{center}
\textbf{\large Supplemental Material for\\[2pt]
``Quasiparticle Theory of Nonlocal Magic in Monitored Fermionic Gaussian Systems''}
\end{center}

\makeatletter
\def\@sectioncntformat#1{\csname the#1\endcsname}%
\@removefromreset{equation}{section}%
\makeatother
\setcounter{secnumdepth}{3}
\setcounter{section}{0}
\setcounter{subsection}{0}
\setcounter{equation}{0}
\setcounter{figure}{0}
\setcounter{table}{0}
\renewcommand{\thesection}{S\arabic{section}}
\renewcommand{\thesubsection}{\Alph{subsection}}
\renewcommand{\theequation}{S\arabic{equation}}
\renewcommand{\thefigure}{S\arabic{figure}}
\renewcommand{\thetable}{S\arabic{table}}

\section{Fermionic nonlocal magic of a two-mode normal form}
\label{sec:supp_twomode}
 
By means of local Gaussian unitaries, any pure fermionic Gaussian state can be brought to a product of two-mode states, each shared between $A$ and $B$ and labeled by an eigenvalue $\nu$ of $i\Gamma_A$~\cite{SuraceTagliacozzo2022,Iannotti2026,collura2026}. After a Jordan--Wigner mapping of the two modes onto qubits, each factor takes the form
\begin{equation}
|\psi_\nu\rangle=\sqrt{\tfrac{1+\nu}{2}}\,|00\rangle+\sqrt{\tfrac{1-\nu}{2}}\,|11\rangle .
\end{equation}
Its only nonvanishing Pauli expectation values are $\langle II\rangle=\langle ZZ\rangle=1$, $\langle ZI\rangle=\langle IZ\rangle=\nu$, and $\langle XX\rangle=-\langle YY\rangle=\sqrt{1-\nu^2}$, so that $\frac14\sum_P\langle P\rangle^4=\frac14[2+2\nu^4+2(1-\nu^2)^2]=1-\nu^2+\nu^4$, and Eq.~\eqref{eq:SRE} gives $M_2(\psi_\nu)=\ms(\nu^2)$. Since the SRE is additive under tensor products, the SRE of the normal form is then $\sum_j\ms(\nu_j^2)$. At R\'enyi index 2, Refs.~\cite{Iannotti2026,collura2026} prove that this value is a local minimum of the SRE over local Gaussian unitaries, and \cite{Iannotti2026} further proves that it is the global minimum within this class of unitaries when the smaller subsystem contains at most three fermionic modes. Following these works, we take this value as the FNLM, Eq.~\eqref{eq:magic}. Note that allowing for arbitrary local unitaries can lower the value further, since the unrestricted nonlocal SRE grows at most logarithmically with the entanglement entropy~\cite{sierant2026,Sierant2026exact}. The fermionic quantity studied here is therefore an upper bound on the unrestricted nonlocal SRE, and the two can scale differently.
 
For number-conserving states, the Majorana covariance is fixed by the correlation matrix $C_A$, with $\nu_j=|1-2\lambda_j|$ and $\lambda_j\in\mathrm{spec}(C_A)$. This gives the weight $\wt(\lambda)=\ms[(1-2\lambda)^2]$ of Eq.~\eqref{eq:weight}, which vanishes at $\lambda=0,\frac12,1$ and reaches its maximum, $\log_2(4/3)$, at $\lambda=(2\mp\sqrt2)/4$. For comparison, the entanglement weight $-\lambda\log_2\lambda-(1-\lambda)\log_2(1-\lambda)$ is maximal precisely at $\lambda=\frac12$~\cite{Peschel2003,PeschelEisler2009}.

\section{Models}
\label{sec:supp_models}

Both models have the form
\begin{equation}
H=\sum_{i,j}h_{ij}d_i^\dagger d_j=\sum_k\epsilon_k d_k^\dagger d_k,\qquad
L_j=\sqrt\gamma\,n_j,\quad n_j=d_j^\dagger d_j ,
\label{eq:supp_models_H}
\end{equation}
where $d_j$ are the monitored modes, and $d_k=L^{-1/2}\sum_je^{-ikj}d_j$. In both models, $H$ conserves the number of $d$ fermions, $N=\sum_jn_j$. The two models differ in the nature of the monitored modes and in the dispersion $\epsilon_k$.

\subsection{Hopping chain}

In the first model, $d_j$ are the physical fermions on the lattice sites, with nearest-neighbour hopping,
\begin{equation}
H_{\rm hop}=-\frac J2\sum_j\big(d_j^\dagger d_{j+1}+d_{j+1}^\dagger d_j\big).
\label{eq:supp_free_H}
\end{equation}
The jumps measure the site occupations, and $N$ is the particle number. For periodic boundary conditions, with $k_m=2\pi m/L$, one finds
\begin{equation}
\epsilon_{\rm hop}(k)=-J\cos k,\qquad v_{\rm hop}(k)=J\sin k .
\label{eq:supp_free_dispersion}
\end{equation}
For open boundary conditions, instead, $H_{\rm hop}$ is diagonalized by the sine transform $S_{jm}=\sqrt{2/(L+1)}\,\sin[\pi m(j+1)/(L+1)]$, with $j=0,\dots,L-1$, $m=1,\dots,L$, and $k_m=\pi m/(L+1)$.

\subsection{Ising-dispersion chain}

In the second model, $d_j$ are the Wannier orbitals of the Bogoliubov quasiparticles of the transverse-field Ising chain,
\begin{equation}
H_{\rm TFIM}=-J\sum_j\sigma_j^x\sigma_{j+1}^x-h\sum_j\sigma_j^z .
\label{eq:supp_tfim_spin}
\end{equation}
After a Jordan--Wigner transformation to fermions $c_j$ and a Fourier transformation, $c_j=L^{-1/2}\sum_ke^{ikj}c_k$, the Hamiltonian reads, in the Nambu basis $\Psi_k=(c_k,c_{-k}^\dagger)^T$ and up to a constant~\cite{SantoroIsing},
\begin{equation}
H_{\rm TFIM}=\frac12\sum_k\Psi_k^\dagger\mathcal H_k\Psi_k,\qquad
\mathcal H_k=A_k\tau^z+B_k\tau^y,\quad A_k=2(h-J\cos k),\quad B_k=2J\sin k ,
\label{eq:supp_tfim_bdg}
\end{equation}
where $\tau^{y,z}$ are Pauli matrices in Nambu space. A Bogoliubov rotation by the angle $\theta_k=\operatorname{atan2}(B_k,A_k)$ gives $H_{\rm TFIM}=\sum_k\epsilon_{\rm I}(k)(\eta_k^\dagger\eta_k-\frac12)$, with
\begin{equation}
\epsilon_{\rm I}(k)=\sqrt{A_k^2+B_k^2}=2\sqrt{(h-J\cos k)^2+J^2\sin^2k},\qquad
v_{\rm I}(k)=\frac{2hJ\sin k}{\sqrt{(h-J\cos k)^2+J^2\sin^2k}} .
\label{eq:supp_tfim_dispersion}
\end{equation}
We then define the monitored modes as $d_j=L^{-1/2}\sum_ke^{ikj}\eta_k$, so that
\begin{equation}
H_{\rm I}=\sum_k\epsilon_{\rm I}(k)\,\eta_k^\dagger\eta_k=\sum_{i,j}h^{\rm I}_{ij}\,d_i^\dagger d_j,\qquad
h^{\rm I}_{ij}=\frac1L\sum_k\epsilon_{\rm I}(k)\,e^{ik(i-j)} ,
\label{eq:supp_bog_realspace}
\end{equation}
a hopping with exponentially decaying amplitudes between Bogoliubov sites away from criticality. The jumps measure the occupations of these sites, and $N=\sum_jd_j^\dagger d_j=\sum_k\eta_k^\dagger\eta_k$ is the number of Bogoliubov quasiparticles, which commutes with $H_{\rm I}$, unlike the number of Jordan--Wigner fermions $\sum_jc_j^\dagger c_j$. Since the modes $d_j$ are nonlocal in the physical spins, this model shares the Ising dispersion but not the physical measurement setup: we use it only to vary the velocity distribution at fixed measurement structure, while the physically monitored Ising chain is analyzed in Sec.~\ref{sec:supp_tfim}. For periodic spin boundary conditions in the even-parity sector, the momenta are $k_m=2\pi(m+\frac12)/L$; for open boundaries, we use $h^{\rm I}=S\,\mathrm{diag}[\epsilon_{\rm I}(k_m)]\,S^\dagger$ with $k_m=\pi m/(L+1)$. At $h=0.8$ and $J=1$, the maximal velocity is $\max_k|v_{\rm I}|=2h=1.6$, reached at $\cos k=h$, to be compared with $J$ for the hopping chain.

\section{Multiplet quasiparticles}
\label{sec:supp_multiplet}

We consider the periodic product state $|1\,\underbrace{0\cdots0}_{p-1}\,1\,\underbrace{0\cdots0}_{p-1}\,\cdots\rangle$, with one particle per unit cell of length $p$ and filling $\rho=1/p$. For this initial state, the correlations are carried by multiplets of $p$ modes rather than by independent pairs~\cite{Bertini2018,BastianelloCollura2020},
\begin{equation}
k_a=k+\frac{2\pi a}{p},\qquad a=0,\dots,p-1,\qquad k\in\Big[-\frac\pi p,\frac\pi p\Big).
\end{equation}

\paragraph{Translation symmetry and multiplets.---}
The multiplet structure is a direct consequence of the symmetry of the initial state. Let $T$ be the translation by one site, $Td_xT^\dagger=d_{x+1}$, so that $T^pd_kT^{-p}=e^{ikp}d_k$. For a number-conserving Gaussian state, the following two properties are equivalent: (a) the state is invariant, up to a phase, under translations by $p$ sites; (b) its momentum correlations $\langle d_k^\dagger d_q\rangle$ vanish unless $q-k\in\frac{2\pi}{p}\mathbb Z$, i.e., each momentum $k$ is correlated only with the members of its multiplet $\{k_a\}$. Indeed, if $T^p|\psi\rangle=e^{i\varphi}|\psi\rangle$, then $\langle d_k^\dagger d_q\rangle=e^{i(q-k)p}\langle d_k^\dagger d_q\rangle$, which implies (b). Conversely, if (b) holds, the correlation matrix $C_{xy}=\frac1L\sum_{k,q}e^{-ikx+iqy}\langle d_k^\dagger d_q\rangle$ satisfies $C_{x+p,y+p}=C_{xy}$, since $e^{i(q-k)p}=1$ for all correlated pairs of momenta; and since a Slater determinant is fixed by its correlation matrix, the state is invariant under translations by $p$ sites. Pairs are the special case $p=2$, with members $k$ and $k+\pi$, which have opposite velocities. The product form with one particle per unit cell adds two further properties: each multiplet carries exactly one fermion, and its members have equal occupations $1/p=\rho$. In particular, among these states, half filling requires $p=2$.

\paragraph{Exact correlation structure.---}
The multiplet structure follows exactly from the block-Toeplitz form of the correlation matrix $C_{xy}(t)=\langle d_x^\dagger d_y\rangle_t$, with $d_x=L^{-1/2}\sum_ke^{ikx}d_k$. For the initial state, $C(0)$ is diagonal, with $C_{xx}(0)=1$ for $x\in p\mathbb Z$ and $0$ otherwise, so that
\begin{equation}
\langle d_k^\dagger d_q\rangle_0=\frac1L\sum_{x\in p\mathbb Z}e^{i(k-q)x}=\frac1p\,\delta_{k-q\in\frac{2\pi}{p}\mathbb Z}.
\label{eq:supp_mom_corr}
\end{equation}
Only the members $k_a=k+2\pi a/p$ of the same multiplet are therefore correlated, with $Q_{ab}(k)=\langle d_{k_a}^\dagger d_{k_b}\rangle_0=1/p$, and each member has occupation $Q_{aa}=1/p=\rho$. Since $H$ is diagonal in momentum, the unitary dynamics gives exactly
\begin{equation}
Q_{ab}(k,t)=\frac1p\,e^{i(\epsilon_{k_a}-\epsilon_{k_b})t},\qquad Q(k,t)=U(k,t)\,Q(k)\,U^\dagger(k,t),
\label{eq:supp_Qt}
\end{equation}
with $U={\rm diag}(e^{i\epsilon_{k_a}t})$: the occupations $1/p$ are conserved, and $Q(k,t)$ remains a rank-one projector, i.e., each multiplet carries exactly one fermion. In real space, writing $x=pX+\alpha$ and $y=pY+\beta$ with $\alpha,\beta=0,\dots,p-1$, the correlation matrix is block Toeplitz,
\begin{equation}
C_{pX+\alpha,\,pY+\beta}(t)=\int_{-\pi}^{\pi}\frac{d\theta}{2\pi}\,e^{-i\theta(X-Y)}\,\mathcal C_{\alpha\beta}(\theta,t),\qquad
\mathcal C(\theta,t)=W(\theta)\,Q(k,t)\,W^\dagger(\theta),
\label{eq:supp_symbol}
\end{equation}
with $k=\theta/p$ and the unitary matrix $W_{\alpha a}(\theta)=e^{-ik_a\alpha}/\sqrt p$. The $p\times p$ symbol $\mathcal C(\theta,t)$ is thus unitarily equivalent to $Q(k,t)$, and its spectrum, $\{1,0,\dots,0\}$, is conserved exactly by the unitary dynamics. Finally, restricting the rank-one matrix $Q(k,t)$ to any subset of $m$ members leaves a single nonzero eigenvalue, $\sum_{a=1}^{m}|e^{i\epsilon_{k_a}t}|^2/p=m/p$, independently of the phases. The only approximation in Eq.~\eqref{eq:multiplet_magic} is the assignment of members to $A$ through their positions $x+v_{k_a}t$, which follows from the stationary-phase asymptotics of the spectrum of the truncated block-Toeplitz matrix $C_A$ and holds at leading order for $t,\ell\to\infty$ at fixed $t/\ell$~\cite{Fag2008,Bertini2018,BastianelloCollura2020}.

\begin{figure}[t]
\centering
\includegraphics[width=0.6\linewidth]{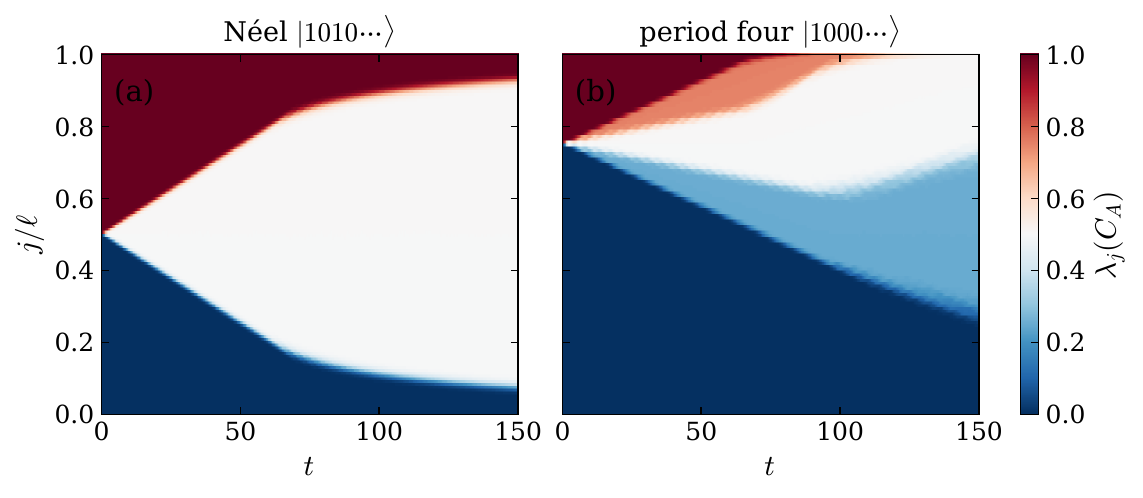}
\caption{Unitary evolution of the eigenvalues $\lambda_j$ of the reduced correlation matrix $C_A$ for (a) the N\'eel state $|1010\cdots\rangle$ and (b) the period-four state $|1000\cdots\rangle$, plotted against the normalized index $j/\ell$. While the N\'eel spectrum accumulates at $\lambda=\frac12$, which carries no FNLM, the period-four spectrum develops bands at $\lambda=\frac14,\frac34$, which carry the weight $\mu_4(1)$. The color scale encodes the eigenvalue; here $L=512$, $\ell=128$, and $J=1$.}
\label{fig:eig_spectra_free}
\end{figure}

\begin{figure}[t]
\centering
\resizebox{0.9\linewidth}{!}{%
\begin{tikzpicture}[
    >=Latex,
    font=\small,
    line cap=round,
    line join=round,
    panel/.style={font=\bfseries},
    box/.style={rounded corners=3pt, draw=black!20, fill=black!3, inner sep=5pt},
    every node/.style={align=center}
]
\colorlet{c0}{blue!75!black}
\colorlet{c1}{teal!70!black}
\colorlet{c2}{orange!85!black}
\colorlet{c3}{red!75!black}
\colorlet{rbzfill}{blue!8}
\colorlet{subfill}{teal!10}
 
\begin{scope}[shift={(0,0)}]
\node[panel] at (0,3.0) {(a) Initial momentum multiplet};
\draw[gray!70] (-2.8,0) -- (2.8,0);
\draw[->,gray!70] (2.8,0) -- (3.1,0) node[right] {$k$};
\fill[rbzfill,rounded corners=2pt] (-2.1,-0.23) rectangle (2.1,0.23);
\draw[dashed,gray!65] (-2,0.55) -- (-2,-0.55);
\draw[dashed,gray!65] ( 2,0.55) -- ( 2,-0.55);
\node[fill=white,inner sep=1pt] at (0,0.42) {RBZ};
\node[below] at (-2,0) {$-\pi/4$};
\node[below] at ( 2,0) {$\pi/4$};
\fill[c0] (0,0) circle (2.3pt);
\node[below=5pt] at (0,0) {$k$};
\fill[c0] (-2.25,-1.45) circle (2.4pt);
\fill[c1] (-0.75,-1.45) circle (2.4pt);
\fill[c2] ( 0.75,-1.45) circle (2.4pt);
\fill[c3] ( 2.25,-1.45) circle (2.4pt);
\node[below=2pt,text=c0] at (-2.25,-1.45) {$k_0$};
\node[below=2pt,text=c1] at (-0.75,-1.45) {$k_1$};
\node[below=2pt,text=c2] at ( 0.75,-1.45) {$k_2$};
\node[below=2pt,text=c3] at ( 2.25,-1.45) {$k_3$};
\draw[->,thick,c0] (0,-0.03) -- (-2.18,-1.23);
\draw[->,thick,c1] (0,-0.03) -- (-0.80,-1.23);
\draw[->,thick,c2] (0,-0.03) -- ( 0.70,-1.23);
\draw[->,thick,c3] (0,-0.03) -- ( 2.18,-1.23);
\node[box,text width=4.8cm] at (0,-4.00) {%
\[
Q_{ab}(k,0)=\frac14
\]
rank-one occupied orbital\\[2pt]
equally distributed over\\[2pt]
\[
k_a=k+\frac{\pi a}{2}
\]
};
\end{scope}
 
\begin{scope}[shift={(8.3,0)}]
\node[panel] at (0,3.0) {(b) Semiclassical multiplet propagation};
\draw[gray!70] (-2.1,0) -- (4.2,0);
\draw[->,gray!70] (4.2,0) -- (4.5,0) node[right] {$x$};
\fill[subfill] (0,-1.95) rectangle (2.0,1.95);
\draw[black!70] (0,-1.95) -- (0,1.95);
\draw[black!70] (2.0,-1.95) -- (2.0,1.95);
\node[fill=white,inner sep=1pt] at (1.0,2.25) {$A=[0,\ell]$};
\fill[black] (-1.55,0) circle (2pt);
\node[below left=1pt] at (-1.55,0) {$x$};
\draw[->,very thick,c0] (-1.55,0) -- (-0.20,1.35);
\draw[->,very thick,c1] (-1.55,0) -- (0.85,0.75);
\draw[->,very thick,c2] (-1.55,0) -- (1.45,0.05);
\draw[->,very thick,c3] (-1.55,0) -- (2.75,-1.10);
\node[text=c0,above left=0pt]  at (-0.20,1.35) {$v_{k_0}$};
\node[text=c1,above left=0pt]  at ( 0.85,0.75) {$v_{k_1}$};
\node[text=c2,above left=0pt]  at ( 1.45,0.05) {$v_{k_2}$};
\node[text=c3,below right=0pt] at ( 2.75,-1.10) {$v_{k_3}$};
\node[box,text width=6.4cm] at (0.95,-3.75) {%
\[
m(k,x,t)=\sum_{a=0}^{3}
\Theta(x+v_{k_a}t)\Theta(\ell-x-v_{k_a}t)
\]
};
\node[box,text width=5.5cm] at (6.75,0.3) {%
\[
\lambda_m=\frac{m}{4},
\qquad
\mu_4(m)=\mathfrak{m}_2\!\left[\left(1-\frac{m}{2}\right)^2\right]
\]
\vspace{2pt}
\begin{tabular}{c c}
$m$ & $\mu_4(m)$ \\ \hline
$1,3$ & $\mathfrak{m}_2(1/4)$ \\
$2$   & $0$ \\
$0,4$ & $0$
\end{tabular}
};
\end{scope}
\end{tikzpicture}%
}
\caption{Multiplet quasiparticle picture for the period-four state.
(a) The initial state $|1000\,1000\cdots\rangle$ correlates the four modes $k_a=k+\pi a/2$, forming a rank-one occupied multiplet.
(b) Under unitary evolution, the four members propagate ballistically with velocities $v_{k_a}$, and the FNLM of a multiplet depends on the number $m$ of its members inside $A=[0,\ell]$; only $m=1$ and $m=3$ contribute.}
\label{fig:multiplet_tikz}
\end{figure}
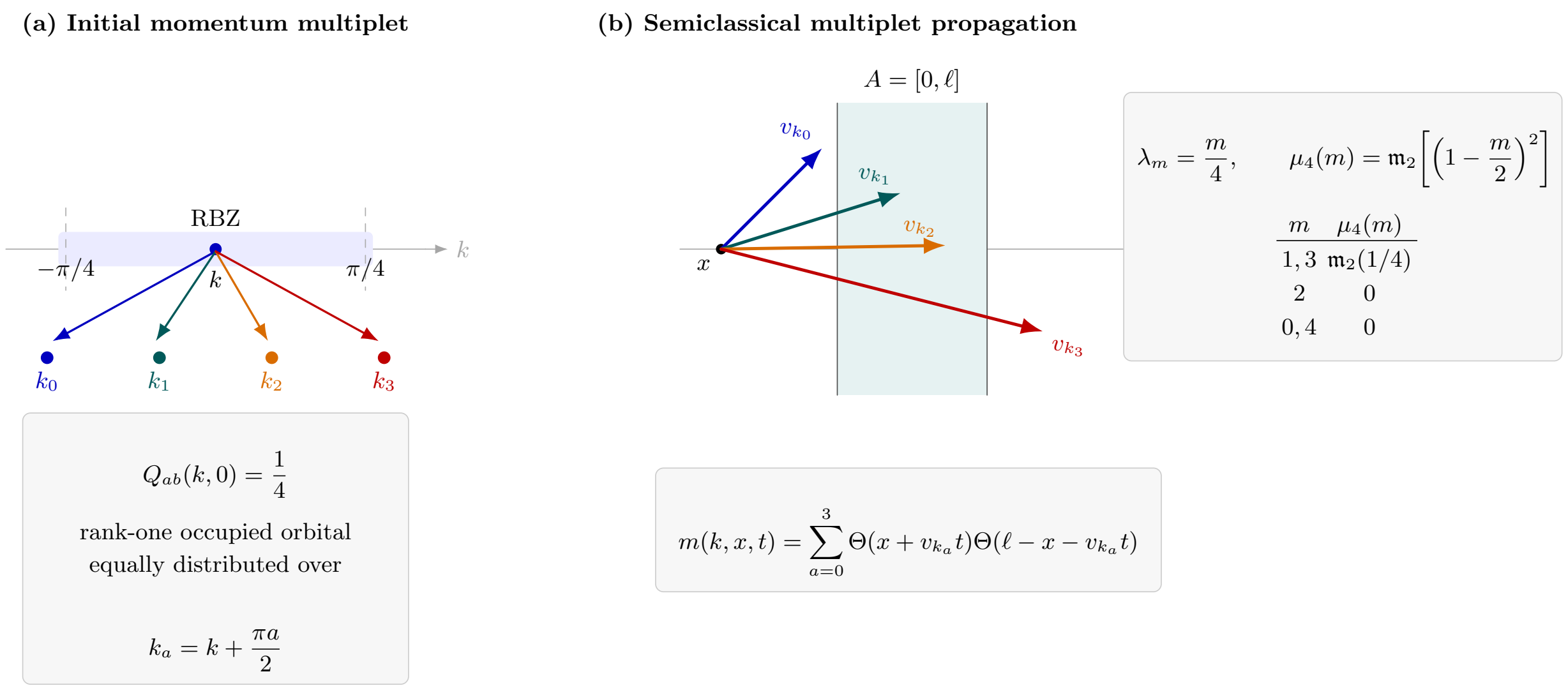

\begin{figure*}[t]
\centering
\includegraphics[width=0.95\textwidth]{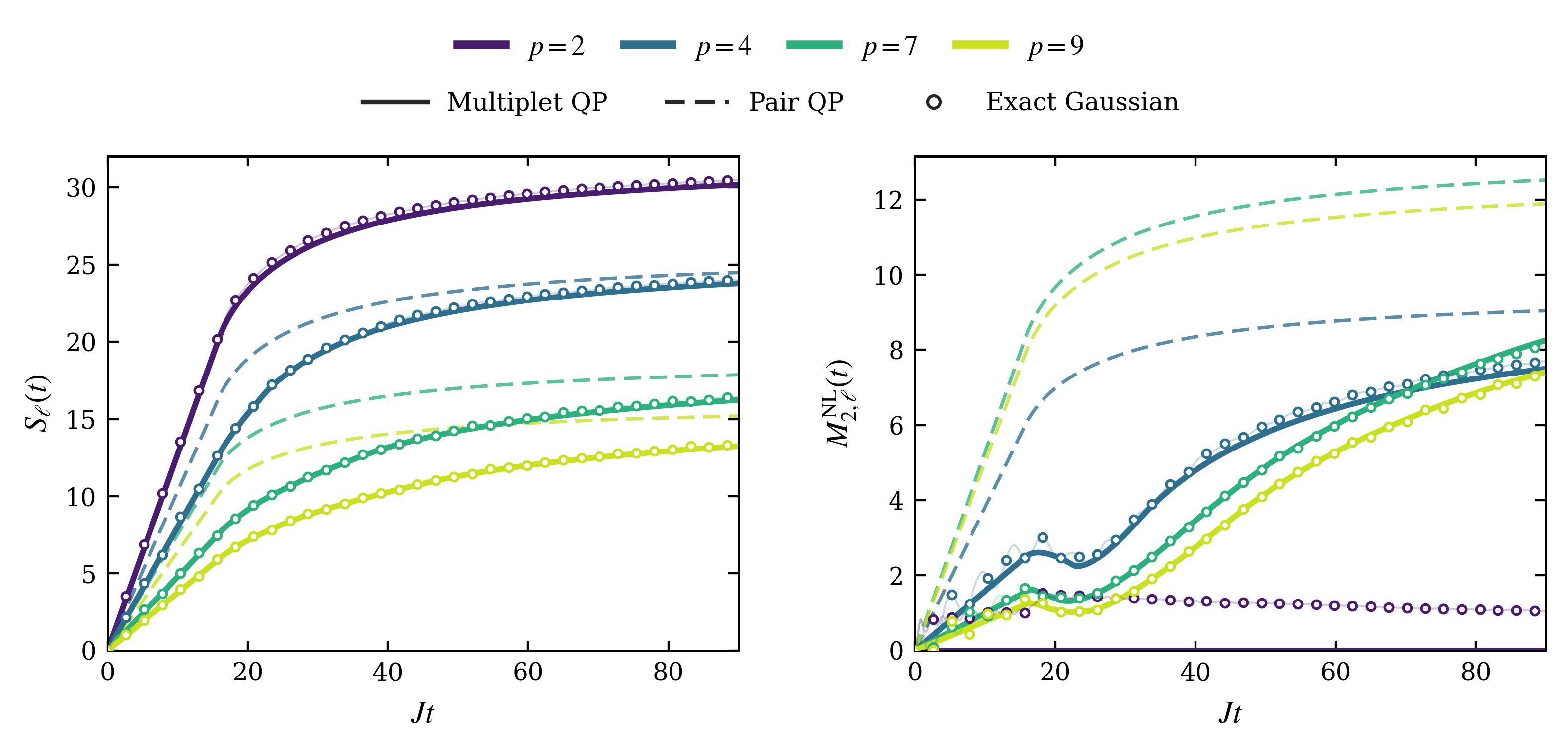}
\caption{Unitary dynamics of (a) the entanglement entropy $S_\ell(t)$ and (b) the FNLM $\Mf_\ell(t)$ following a quench from $|1\,\protect\underbrace{0\cdots0}_{p-1}\,1\,\protect\underbrace{0\cdots0}_{p-1}\,\cdots\rangle$, for $p=2,4,7,9$ (different colors), in the hopping chain. Open symbols represent exact Gaussian dynamics, solid lines the multiplet prediction, Eq.~\eqref{eq:multiplet_magic}, and dashed lines the conventional pair approximation. We fix $J=1$ and $\ell=32$, with $L=512,512,511,513$ for $p=2,4,7,9$, respectively. For $p=2$ the leading quasiparticle contribution to the FNLM vanishes, whereas the exact dynamics retains a subleading contribution.}
\label{fig:unitary_multiplet}
\end{figure*}

Under unitary evolution, the members propagate ballistically with velocities $v_{k_a}$, and the number of members of a multiplet emitted at $x$ that lie in $A=[0,\ell]$ at time $t$ is
\begin{equation}
m(k,x,t)=\sum_{a=0}^{p-1}\Theta(x+v_{k_a}t)\,\Theta(\ell-x-v_{k_a}t).
\label{eq:multiplet_count}
\end{equation}
Each multiplet then contributes the weight
\begin{equation}
\mu_p(m)=\wt(m/p)=\ms\Big[\Big(1-\frac{2m}{p}\Big)^2\Big].
\label{eq:multiplet_weight}
\end{equation}
Integrating over the emission points and over the reduced Brillouin zone, we obtain
\begin{equation}
\Mf^{\rm mult}_\ell(t)=\int_{-\pi/p}^{\pi/p}\frac{dk}{2\pi}\int_{-\infty}^{\infty}dx\;\mu_p[m(k,x,t)] .
\label{eq:multiplet_magic}
\end{equation}
Here the $k$-integral counts each multiplet once and equals $1/p$, i.e., the density of particles. Only multiplets shared between $A$ and its complement contribute, since $\mu_p(0)=\mu_p(p)=0$. For $p=4$, one has $\mu_4(1)=\mu_4(3)=\ms(1/4)=\log_2(16/13)$ and $\mu_4(2)=0$, whereas for $p=2$ the only nontrivial partition has $\lambda=\frac12$, so that the leading contribution vanishes. Figure~\ref{fig:eig_spectra_free} shows the resulting spectra: the N\'eel spectrum accumulates at $\lambda=\frac12$, while the period-four spectrum develops bands at $\lambda=\frac14,\frac34$. At long times, each multiplet has at most one member in $A$; since the density of members is $p\times(1/p)=1$ per site, this gives
\begin{equation}
\lim_{t\to\infty}\Mf^{\rm mult}_\ell(t)=\ell\,\mu_p(1)=\ell\,\wt(\rho).
\label{eq:supp_unitary_sat}
\end{equation}
Note that the approach to this value is slow, because members with nearly equal velocities, close to the edges of the reduced zone, separate slowly. The construction is illustrated in Fig.~\ref{fig:multiplet_tikz}. In Fig.~\ref{fig:unitary_multiplet}, we compare Eq.~\eqref{eq:multiplet_magic} with the exact Gaussian dynamics for $p=2,4,7,9$: the multiplet theory captures the leading ballistic behavior of both entanglement and FNLM, whereas the pair approximation fails for $p>2$. For $p=4$ and $\ell=32$, the multiplet prediction for the FNLM decreases from $2.61$ at $t\simeq17.7$ to $2.25$ at $t\simeq23.2$, before saturating to $\ell\mu_4(1)$. In this window, the fastest members leave $A$, and multiplets with three members inside (weight $\mu_4(1)$) turn into multiplets with two members inside (zero weight). For the entanglement entropy, instead, two members inside carry the maximal weight, so that the prediction grows monotonically. The exact data display the same dip in the FNLM, which is therefore a direct signature of the multiplet structure and of the different weights of entanglement and nonlocal magic.

\paragraph{General multiplets and inhomogeneous states.---}
The eigenvalue of a restricted multiplet does not rely on equal weights. If a multiplet carries one fermion in the orbital $\sum_a\psi_a|k_a\rangle$, its correlation matrix $Q_{ab}=\psi_a^*\psi_b$ is rank one, and its restriction to the members inside $A$ has the single nonzero eigenvalue $\sum_{a\in A}|\psi_a|^2$, independently of the phases acquired under the unitary dynamics; for the periodic states, $|\psi_a|^2=1/p$ and the eigenvalue is $m/p$. Moreover, at leading order for $\ell,t\gg p$, the correlations emitted from a point $x$ are determined by the initial state within a distance of order $p$ of $x$~\cite{Bertini_2018,Alba2018}. Equation~\eqref{eq:multiplet_magic} therefore extends to initial states that are periodic only locally: the $x$-integral runs over the emission points, $p$, the RBZ, and $\mu_p$ are those of the initial state at $x$, and empty regions emit nothing. A shift of the occupation pattern by $s$ sites multiplies the momentum components of each multiplet by phases, $Q_{ab}=\frac1pe^{i(k_a-k_b)s}$, and leaves its restricted eigenvalue unchanged. Regions of size $O(p)$ around interfaces contribute only subleading corrections.

\section{Quantum-jump dynamics and kinetic equation}
\label{sec:supp_kinetic}
 
In this section, we derive the kinetic equation for the monitored models from the quantum-jump unraveling.
 
\paragraph{Stochastic dynamics.---}
The normalized state obeys the stochastic Schr\"odinger equation~\cite{DalibardCastinMolmer1992,PlenioKnight1998,Daley2014}
\begin{equation}
d|\psi_t\rangle=\Big[-iH-\frac12\sum_j\big(L_j^\dagger L_j-\langle L_j^\dagger L_j\rangle_t\big)\Big]|\psi_t\rangle\,dt
+\sum_j\Big(\frac{L_j}{\sqrt{\langle L_j^\dagger L_j\rangle_t}}-1\Big)|\psi_t\rangle\,dN_j,
\label{eq:supp_jump_sse}
\end{equation}
where $dN_j\in\{0,1\}$ are Poisson increments with $\mathbb E[dN_j|\psi_t]=\gamma\langle n_j\rangle_t\,dt$. Since $n_j^2=n_j$ and $N$ is conserved, one has $\sum_j(L_j^\dagger L_j-\langle L_j^\dagger L_j\rangle)=\gamma(N-\langle N\rangle)=0$ on states of fixed particle number. As a consequence, the evolution between jumps is unitary, and jumps occur at the total rate $R=\gamma N$.
 
\paragraph{Covariance update.---}
For a conditional Slater determinant, $C_{ij}=\langle d_i^\dagger d_j\rangle$ is the projector onto the occupied subspace, $C^\dagger=C=C^2$, and between jumps it evolves as $\dot C=-i[h,C]$. A jump at site $j$ occurs with probability proportional to $p_j=C_{jj}$ and maps $|\psi\rangle\to n_j|\psi\rangle/\sqrt{p_j}$. Denoting by $P_j=|j\rangle\langle j|$ the single-particle projector onto site $j$, the corresponding covariance update reads
\begin{equation}
C^{(j)}=C+P_j-\frac{CP_jC}{p_j}.
\label{eq:supp_jump_map}
\end{equation}
This result can be understood by noting that the jump removes the normalized occupied orbital $C|j\rangle/\sqrt{p_j}$ and adds the localized orbital $|j\rangle$, while the remaining occupied subspace is unchanged and orthogonal to $|j\rangle$: indeed, any $|\phi\rangle$ in the range of $C$ satisfies $\langle\phi|j\rangle=\langle\phi|C|j\rangle$, so that orthogonality to $C|j\rangle$ implies orthogonality to $|j\rangle$. Each jump therefore renews exactly one occupied orbital. Since jumps occur at total rate $\gamma N$, each of the $N$ orbitals is renewed at the average rate $\gamma$, independently of the filling, which fixes the collapse rate in Eq.~\eqref{eq:renewal}. For one particle per unit cell, each multiplet of Sec.~\ref{sec:supp_multiplet} is one occupied orbital; postulate (i) of the main text takes it to survive as a whole with probability $e^{-\gamma t}$.

\begin{figure}[htbp]
\centering
\includegraphics[width=0.6\linewidth]{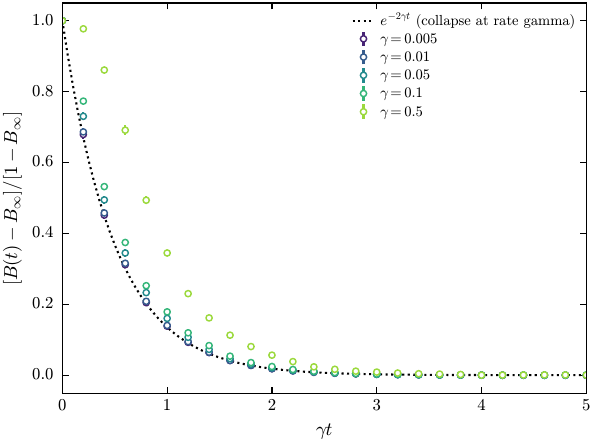}
\caption{Weight of the multiplet correlations, Eq.~\eqref{eq:supp_block_weight}, as $[B(t)-B_\infty]/[1-B_\infty]$ against $\gamma t$, for the period-four state in the hopping chain with $L=512$ and $N_{\rm traj}=200$ trajectories; $B_\infty$ is the average over $t\geq6/\gamma$. The line is $e^{-2\gamma t}$, which corresponds to correlations decaying at the rate $\gamma$.}
\label{fig:multiplet_survival}
\end{figure}

\paragraph{Decay of the multiplet correlations.---}
The collapse rate entering postulate (i) can be tested directly on the trajectories, independently of the FNLM. For the period-$p$ state, the correlations carried by the multiplet with reduced momentum $k$ are $Q_{ab}(k,t)=\langle d^\dagger_{k_a}d_{k_b}\rangle_t$, with $a,b=0,\dots,p-1$, and we measure their total weight
\begin{equation}
B(t)=\frac1N\sum_{k\in{\rm RBZ}}{\rm Tr}\,\big[Q(k,t)\,Q^\dagger(k,t)\big]
\label{eq:supp_block_weight}
\end{equation}
on each trajectory. Without jumps, $B(t)=1$ at all times, since each $Q(k,t)$ remains a rank-one projector [Eq.~\eqref{eq:supp_Qt}] and the unitary dynamics only adds phases. Any decay of $B$ is therefore due to the jumps. A jump at site $j$ removes the orbital $C|j\rangle/\sqrt{\langle n_j\rangle}$, which overlaps every multiplet orbital with the same weight $p/L=1/N$, and replaces it with the localized orbital $|j\rangle$. At the total jump rate $\gamma N$, the amplitude of each multiplet thus decays as $e^{-\gamma t}$, and $B(t)$, which is quadratic in this amplitude, as $e^{-2\gamma t}$, above the background $B_\infty$ left by the renewed orbitals. Figure~\ref{fig:multiplet_survival} shows that $[B(t)-B_\infty]/[1-B_\infty]$ collapses onto $e^{-2\gamma t}$ for all monitoring rates, with no free parameters, which confirms that the initial correlations decay at the rate $\gamma$ assumed in Eq.~\eqref{eq:hybrid}. Two limitations should be noted. First, since $B$ sums over all emission points, it does not resolve where in space a jump destroys correlations; this locality is tested by the FNLM itself, through the factor $e^{-\gamma t}$ in Eq.~\eqref{eq:hybrid}. Second, $B$ probes only the decay of the initial correlations and not the structure of the renewed ones, which postulates (ii) and (iii) describe and which are tested by the stationary data (Fig.~\ref{fig:scaling_analysis} and Table~\ref{tab:weights}).

\paragraph{Averaged dynamics.---}
Combining the unitary evolution and the stochastic jumps, we obtain the trajectory-level equation
\begin{equation}
dC=-i[h,C]\,dt+\sum_j\Big(P_j-\frac{CP_jC}{p_j}\Big)dN_j .
\label{eq:supp_jump_cov_sde}
\end{equation}
Averaging with $\mathbb E[dN_j|C]=\gamma p_j\,dt$, and using $\sum_jp_jP_j=\mathrm{diag}(C)$ and $\sum_jCP_jC=C^2=C$, we arrive at the closed linear equation
\begin{equation}
\partial_t\avg C=-i[h,\avg C]-\gamma\big[\avg C-\mathrm{diag}(\avg C)\big],
\label{eq:supp_jump_meanC}
\end{equation}
where $\avg C=\mathbb E[C]$. This is the correlation-matrix equation of the dephasing Lindbladian~\cite{Eisler2011,Znidaric2010,BernardJin2019}, and its closure relies on the purity $C^2=C$ of each trajectory. We stress that Eq.~\eqref{eq:supp_jump_meanC} fixes the averaged two-point function, but not trajectory averages of nonlinear functionals such as the FNLM.
 
\paragraph{Kinetic equation.---}
To obtain a large-scale description, we introduce the Wigner function of $\avg C$,
\begin{equation}
n(x,k,t)=\sum_re^{-ikr}\,\avg C_{x+r/2,\,x-r/2}(t),
\label{eq:supp_jump_wigner}
\end{equation}
where the sum runs over the relative coordinates compatible with lattice sites, and we treat $x$ as a continuous variable on scales much larger than the lattice spacing~\cite{Fagotti2017}. To leading order in gradients, the unitary term becomes $-i[h,\avg C]\to-v_k\partial_xn$, while the measurement term, which is local, gives $-\gamma[n-\rho(x,t)]$ with $\rho=\int\frac{dk}{2\pi}n$. Hence
\begin{equation}
\partial_tn(x,k,t)=-v_k\partial_xn(x,k,t)-\gamma\Big[n(x,k,t)-\int_{-\pi}^{\pi}\frac{dq}{2\pi}\,n(x,q,t)\Big].
\label{eq:supp_jump_ghd}
\end{equation}
This is a relaxation-time (Boltzmann-type) kinetic equation, which coincides with the one obtained for diffusive occupation monitoring~\cite{Cao2019}. Integrating over $k$ yields the continuity equation $\partial_t\rho+\partial_xj=0$, with $j=\int\frac{dk}{2\pi}v_kn$.

\paragraph{Renewal weight.---}
In postulate (iii) of the main text, a renewed pair carries the weight $\ms[\rho(x,s)]$ of the local filling at its emission point and time. For a product state in the occupation basis, $\avg C(0)$ is diagonal, so that $n(x,k,0)=\langle n_x\rangle$ for all $k$, which after coarse-graining over a unit cell gives $n(x,k,0)=\rho$. Such a homogeneous and $k$-independent occupation is stationary under Eq.~\eqref{eq:supp_jump_ghd}, since the drift term vanishes for $\partial_xn=0$ and the relaxation term vanishes for $n=\rho$. The filling therefore remains constant, and so does the renewal weight $\ms(\rho)$, which yields Eq.~\eqref{eq:hybrid}.
For an inhomogeneous initial state, $n(x,k,0)=\rho_0(x)$ is still $k$-independent, with $\rho_0$ the coarse-grained initial filling, but it is no longer stationary: particles spread ballistically and relax at rate $\gamma$, and we obtain $\rho(x,t)$ by integrating Eq.~\eqref{eq:supp_jump_ghd} numerically.

Since the weight is evaluated at the trajectory-averaged occupation, while the FNLM is computed from the fluctuating eigenvalues of each trajectory, this prescription is an approximation; Sec.~\ref{sec:supp_halffilling} shows that it fails at half filling.

\section{Stationary limit and finite rings}
\label{sec:supp_stationary}
 
At long times, the multiplet term of Eq.~\eqref{eq:hybrid} decays as $e^{-\gamma t}$, and the renewal term tends to
\begin{equation}
\avg{\Mf}^{\rm st}_\ell=\ms(\rho)\int_{-\pi}^{\pi}\frac{dk}{2\pi}\int_0^\infty da\,\gamma e^{-\gamma a}\min(2|v_k|a,\ell)
=\ms(\rho)\int_{-\pi}^{\pi}\frac{dk}{2\pi}\,\frac{2|v_k|}{\gamma}\Big(1-e^{-\gamma\ell/2|v_k|}\Big),
\end{equation}
which gives Eqs.~\eqref{eq:scaling} and \eqref{eq:fM}. For $\gamma\ell\ll|v_k|$, the bracket expands to $\gamma\ell/2|v_k|$ and one recovers $\ell\,\ms(\rho)$, while for $\gamma\ell\gg|v_k|$ it tends to $1$, giving the area law $\ms(\rho)\gamma^{-1}\int\frac{dk}{2\pi}2|v_k|$. For the hopping chain, $\int\frac{dk}{2\pi}2|v_k|=4J/\pi$.
 
On a ring of $L$ sites, a pair with separation $d=2|v_k|a$ contributes whenever exactly one of its members lies in $A$. Defining $d_L=\min(d\bmod L,\,L-d\bmod L)$, the factor $\min(2|v_k|a,\ell)$ in Eq.~\eqref{eq:hybrid} becomes
\begin{equation}
\mathcal L^{(L)}_k(a)=\min\big(d_L,\;\ell,\;L-\ell\big),
\label{eq:supp_ring}
\end{equation}
which respects the pure-state symmetry $\ell\to L-\ell$, and the analogous replacement in Eq.~\eqref{eq:multiplet_magic} counts member positions modulo $L$.

\begin{figure}[htbp]
\centering
\includegraphics[width=0.85\linewidth]{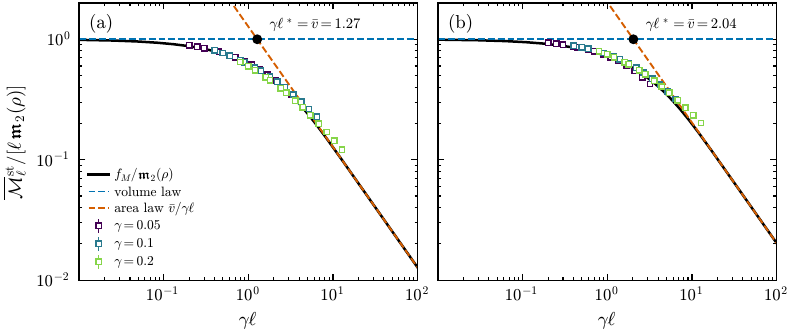}
\caption{Stationary FNLM per site, normalized by the renewal weight, $\avg{\Mf}^{\rm st}_\ell/[\ell\,\ms(\rho)]$, against $\gamma\ell$ for the period-four state in (a) the hopping chain and (b) the Ising-dispersion chain ($h=0.8$), with $L=256$ and $\gamma=0.05,0.1,0.2$ (symbols). The solid line shows $f_M/\ms(\rho)$, Eq.~\eqref{eq:fM}; dashed lines show the volume-law limit $1$ and the area-law limit $\bar v/\gamma\ell$, which meet at $\gamma\ell^*=\bar v$ (dot). Stationary averages are taken over $t\in[5/\gamma,10/\gamma]$ and $N_{\rm traj}=200$ trajectories.}
\label{fig:lstar}
\end{figure}

The volume and area laws meet at $\gamma\ell^*=\bar v$, with $\bar v=\int\frac{dk}{2\pi}2|v_k|$, which gives $\bar v=4J/\pi\simeq1.27J$ for the hopping chain and $\bar v\simeq2.04J$ for the Ising-dispersion chain. Figure~\ref{fig:lstar} shows that the quantum-jump data for different $\gamma$ collapse onto $f_M(\gamma\ell)$ and cross over between the two limits around $\gamma\ell^*$.

\section{Half filling}
\label{sec:supp_halffilling}

\begin{figure}[htbp]
\centering
\includegraphics[width=0.6\linewidth]{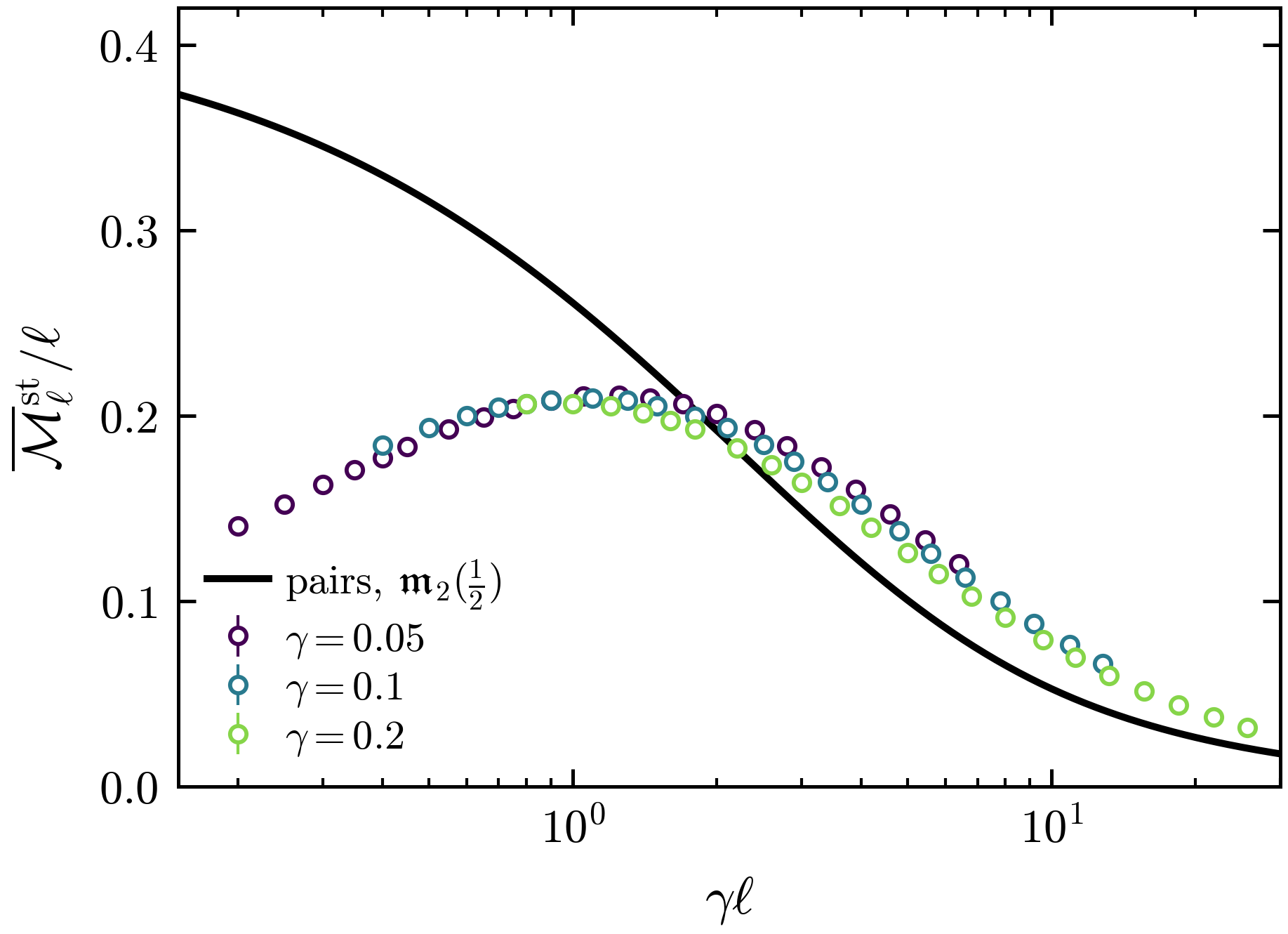}
\caption{Stationary FNLM per site of the N\'eel state against $\gamma\ell$ in the monitored hopping chain, with $\gamma=0.05,0.1,0.2$ and $L=512$. The line shows Eq.~\eqref{eq:fM} with the maximal pair weight $\ms(\frac12)=\log_2(4/3)$; the shaded region above it cannot be reached by any pair weight. Stationary averages are taken over $t\in[5/\gamma,10/\gamma]$ and $N_{\rm traj}=200$ trajectories, with error bars denoting the standard error.}
\label{fig:neel}
\end{figure}

At half filling, Eq.~\eqref{eq:hybrid} predicts $\avg{\Mf}^{\rm st}_\ell=\ell f_M(\gamma\ell)$ with $\ms(\frac12)=\log_2(4/3)$, the largest weight a pair can carry, since $\ms(x)\leq\ms(\frac12)$ for all $x\in[0,1]$. The renewal term of Eq.~\eqref{eq:hybrid} therefore cannot exceed $\log_2(4/3)$ times its geometric factor, whatever weight is assigned to the pairs. Figure~\ref{fig:neel} shows the stationary FNLM per site of the N\'eel state. The data for different $\gamma$ collapse onto a function of $\gamma\ell$, so that the crossover length scales as $\ell^*\propto1/\gamma$, as at other fillings. However, they lie below the prediction at small $\gamma\ell$ and above the prediction at larger $\gamma\ell$, which no pair weight can reach.

\section{Other periods}
\label{sec:supp_periods}

\begin{figure*}[t]
\centering
\includegraphics[width=0.95\textwidth]{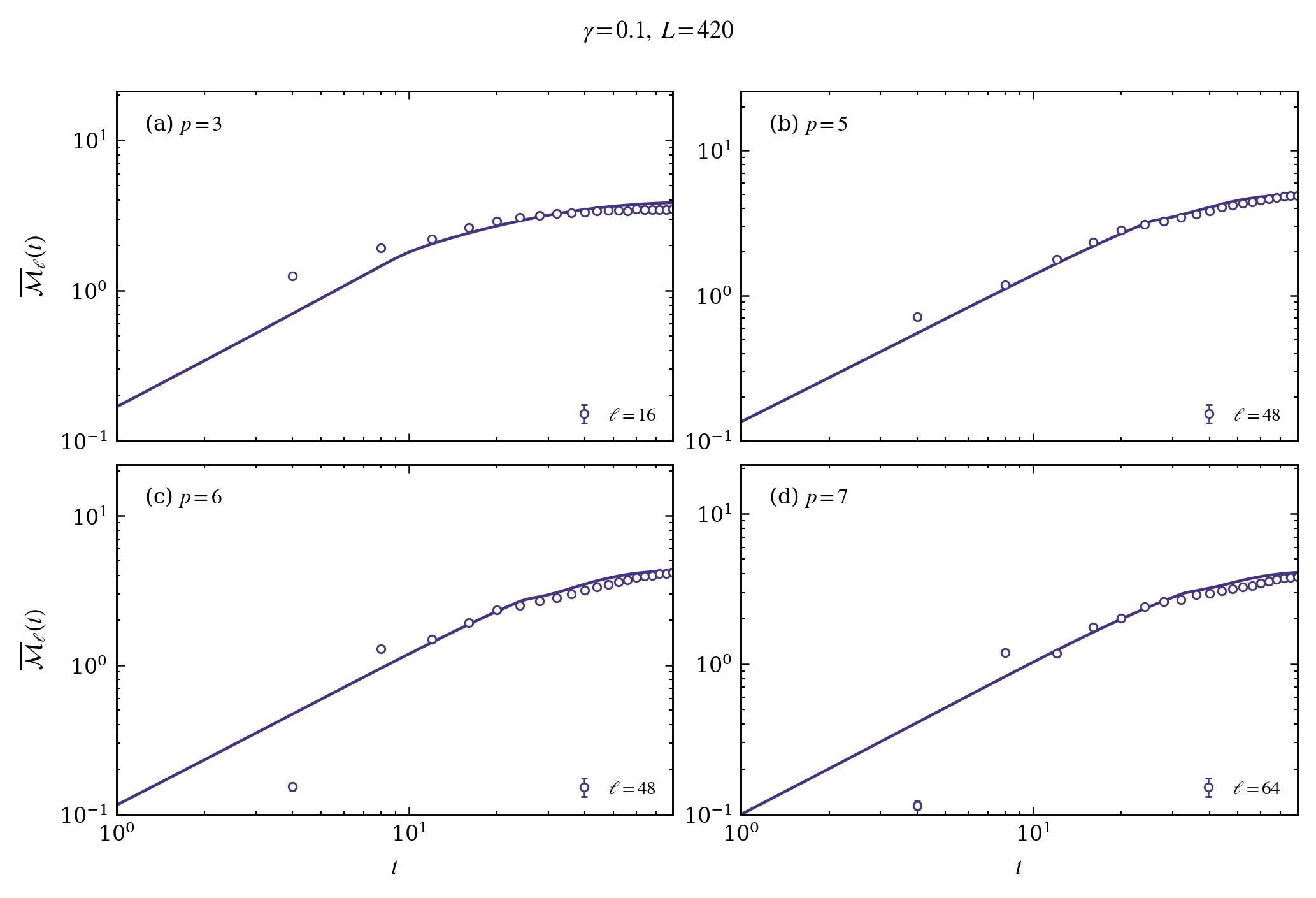}
\caption{Trajectory-averaged FNLM after quenches from  $|1\,\protect\underbrace{0\cdots0}_{p-1}\,1\,\protect\underbrace{0\cdots0}_{p-1}\,\cdots\rangle$, with $p=3,5,6,7$ (left to right) in the hopping chain, for $\gamma=0.1$ and $L=420$, with $\ell=16,48,48,64$, respectively. Symbols: quantum-jump averages with standard errors over $N_{\rm traj}=200$ trajectories; lines: Eq.~\eqref{eq:hybrid} with renewal weight $\ms(1/p)$ and multiplet weights $\mu_p(m)=\wt(m/p)$.}
\label{fig:periods_g01}
\end{figure*}

\begin{figure*}[t]
\centering
\includegraphics[width=0.95\textwidth]{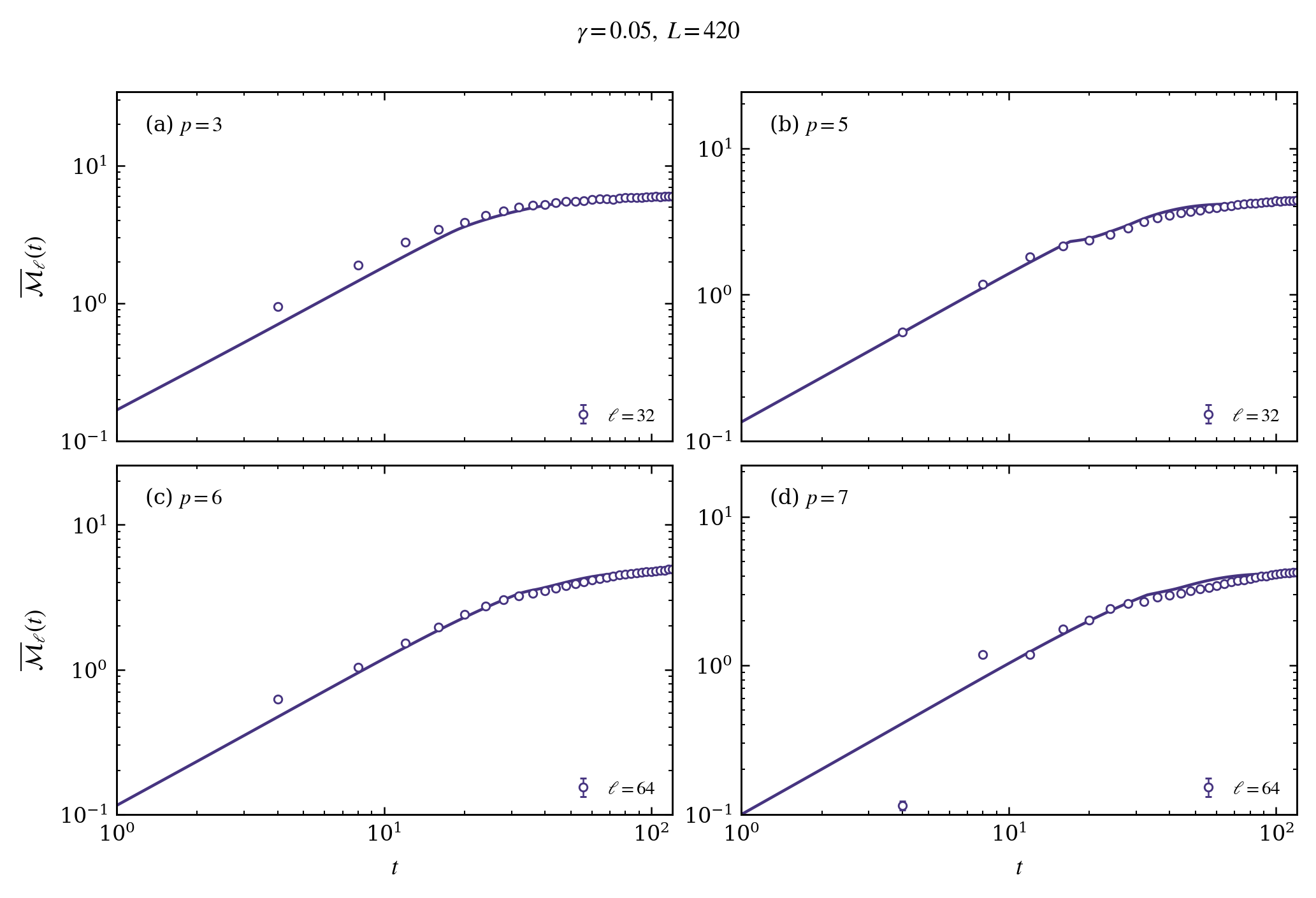}
\caption{Same as Fig.~\ref{fig:periods_g01}, for $\gamma=0.05$ and $L=420$, with $\ell=32,32,64,64$ for $p=3,5,6,7$, respectively.}
\label{fig:periods_g005}
\end{figure*}

We test the theory for the product states $|1\,\underbrace{0\cdots0}_{p-1}\,1\,\underbrace{0\cdots0}_{p-1}\,\cdots\rangle$ with $p=3,5,6,7$, i.e., fillings $\rho=1/p$, at two monitoring rates (Figs.~\ref{fig:periods_g01} and \ref{fig:periods_g005}). The multiplet term is given by Eq.~\eqref{eq:multiplet_magic} with weights $\mu_p(m)=\wt(m/p)$, and the renewal term by Eq.~\eqref{eq:hybrid} with weight $\ms(1/p)$. For all periods, the theory captures the growth and saturation of the FNLM with no free parameters. Note that, for $p\neq4$, the renewal weight $\ms(1/p)$ differs from the unitary saturation density $\mu_p(1)=\wt(1/p)$, Eq.~\eqref{eq:supp_unitary_sat}: the limits $t\to\infty$ and $\gamma\to0$ do not commute, since any nonzero monitoring eventually replaces all initial correlations by renewed ones.

Table~\ref{tab:weights} compares $\wt(1/p)$ and $\ms(1/p)$ with the weight $\mu_{\rm eff}$ extracted from the quantum-jump data, defined as the least-squares constant $\mu$ in $\avg{\Mf}_\ell-e^{-\gamma t}\Mf^{\rm mult}_\ell=\mu\,G_\gamma(t,\ell)$, where $G_\gamma(t,\ell)=\gamma\int_0^t\!da\,e^{-\gamma a}\int\frac{dk}{2\pi}\min(2|v_k|a,\ell)$ is the pair geometry of Eq.~\eqref{eq:hybrid}. The fit uses $\ell\geq4p$, $t\geq20$, and $\gamma=0.05,0.1$.

\begin{table}[htbp]
\centering
\caption{Unitary saturation density $\mu_p(1)=\wt(1/p)$, renewal weight $\ms(1/p)$, and weight $\mu_{\rm eff}$ extracted from the quantum-jump data. The uncertainty is the standard deviation of $[\avg{\Mf}_\ell-e^{-\gamma t}\Mf^{\rm mult}_\ell]/G_\gamma(t,\ell)$ over the fitting window.}
\label{tab:weights}
\begin{tabular}{c c c c c}
\hline\hline
$p$ & $\rho$ & $\wt(\rho)$ & $\ms(\rho)$ & $\mu_{\rm eff}$\\
\hline
3 & $1/3$ & 0.150 & 0.363 & $0.383\pm0.032$ \\
4 & $1/4$ & 0.300 & 0.300 & $0.308\pm0.017$ \\
5 & $1/5$ & 0.378 & 0.252 & $0.253\pm0.013$ \\
6 & $1/6$ & 0.409 & 0.216 & $0.211\pm0.013$ \\
7 & $1/7$ & 0.415 & 0.188 & $0.178\pm0.012$ \\
8 & $1/8$ & 0.408 & 0.167 & $0.154\pm0.010$ \\
9 & $1/9$ & 0.394 & 0.150 & $0.134\pm0.006$ \\
\hline\hline
\end{tabular}
\end{table}
The extracted weight follows $\ms(1/p)$ with a mean absolute percentage error of $\sim 4.95 \%$ (within $\sim10\%$ across all $p$), showing a systematic decrease below it at low filling, whereas it differs strongly from $\wt(1/p)$ for $p\neq4$. Since the quasiparticle description holds in the scaling limit $p\ll\ell\ll L$, with $\ell^*\gg p$, these deviations are expected to include finite-size corrections, which are largest at low filling, where the unit cell is longest.

\section{Inhomogeneous quenches}
\label{sec:supp_junction}
 
\begin{figure*}[t]
\centering
\includegraphics[width=\textwidth]{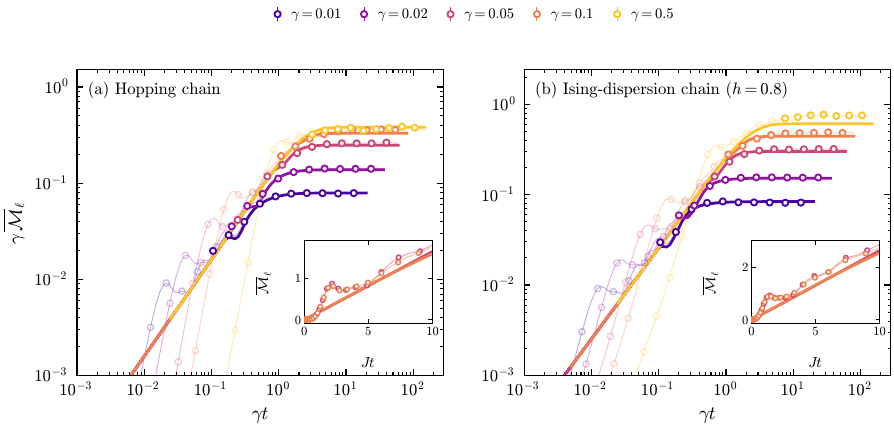}
\caption{Rescaled nonlocal magic $\gamma\avg{\Mf}_\ell$ as a function of $\gamma t$ after the two-reservoir quench, with open boundaries, $L=512$, $\ell=32$, and $\gamma=0.01,0.02,0.05,0.1,0.5$ (from dark purple to yellow). Symbols and lines as in Fig.~\ref{fig:four_models_magic}, with $N_{\rm traj}=200$; (a) hopping chain, (b) Ising-dispersion chain ($h=0.8$).}
\label{fig:junction}
\end{figure*}

In addition to homogeneous quenches, we consider a two-reservoir quench on an open chain whose left and right halves are prepared in the product states $|1000\,1000\cdots\rangle$ and $|0100\,0100\cdots\rangle$, respectively. Both halves have filling $\frac14$, but their occupation patterns are shifted by one site at the junction. The system then evolves under the full Hamiltonian, and we compute the FNLM of the interval $A=\{L/2,\dots,L/2+\ell-1\}$, located immediately to the right of the junction. By Sec.~\ref{sec:supp_multiplet}, the multiplet term is unchanged away from the junction, and the junction contributes only at subleading order. Since both halves also have filling $\frac14$, the coarse-grained Wigner function is $n=\frac14$ on both sides, which is stationary under Eq.~\eqref{eq:supp_jump_ghd}; the renewal weight $\ms(\frac14)$ is therefore uniform and constant. Equation~\eqref{eq:hybrid} thus applies unchanged at leading order, and, as shown in Fig.~\ref{fig:junction}, it describes the data for both dispersions. Note also that, since $A$ lies at a distance $L/2$ from the open boundaries, boundary effects are absent for $t\lesssim L/(2v_{\max})$.
 
\paragraph{Domain wall.---}
For the state of Fig.~\ref{fig:domain_wall}, with $A$ the right half $x>0$, the multiplet term reads
\begin{equation}
\Mf^{\rm mult}(t)=\int_{-\infty}^0dx\int_{\rm RBZ}\frac{dk}{2\pi}\;\mu_4[m(k,x,t)],\qquad m(k,x,t)=\sum_a\Theta(x+v_{k_a}t),
\end{equation}
which grows linearly in $t$. Under monitoring, the renewal term of Eq.~\eqref{eq:renewal} becomes
\begin{equation}
\gamma\int_0^t\!da\,e^{-\gamma a}\int_{v_k>0}\frac{dk}{2\pi}\int_{-v_ka}^{v_ka}dx\;\ms[\rho(x,t-a)],
\end{equation}
with $\rho(x,t)$ obtained from Eq.~\eqref{eq:supp_jump_ghd} with $n(x,k,0)=\frac14\,\Theta(-x)$. Since $A$ lies at a distance $L/2$ from the open boundaries, boundary effects are absent for $t\lesssim L/(2v_{\max})$.

\section{Ising chain with physical occupation monitoring}
\label{sec:supp_tfim}
The no-click quasiparticle theory for this model is presented in Appendix~\ref{app:noclick} of the main text; here we derive the pair dynamics behind Eqs.~\eqref{eq:tfim_complex}--\eqref{eq:tfim_velocity} and discuss the stationary state and the dynamics with quantum jumps.
\paragraph{No-click pair dynamics.---}
For $k>0$, the even-parity pair sector is spanned by $|0_k\rangle$ and $|2_k\rangle=c_k^\dagger c_{-k}^\dagger|0_k\rangle$. With the conventions of Eq.~\eqref{eq:supp_tfim_bdg}, the Hamiltonian in the basis $(|2_k\rangle,|0_k\rangle)$ reads $\mathsf h_k=A_k\sigma^z-B_k\sigma^y$, with eigenvalues $\pm\epsilon_{\rm I}(k)$. Using $c_k^\dagger c_k+c_{-k}^\dagger c_{-k}=(\sigma^z+1)$ in this sector, we find
\begin{equation}
\mathsf h^{\rm eff}_k=\mathcal A_k\sigma^z-B_k\sigma^y-\frac{i\gamma}{2},\qquad \mathcal A_k=A_k-\frac{i\gamma}{2},
\end{equation}
whose eigenvalues are $-\frac{i\gamma}{2}\pm\Omega_k$, with
\begin{equation}
\Omega_k=\sqrt{\mathcal A_k^2+B_k^2}=\sqrt{\epsilon_{\rm I}(k)^2-\frac{\gamma^2}{4}-2i\gamma(h-J\cos k)} ,
\end{equation}
i.e., Eq.~\eqref{eq:tfim_complex}. Since $(\mathcal A_k\sigma^z-B_k\sigma^y)^2=\Omega_k^2\mathbb 1$, the normalized pair state can be written in closed form,
\begin{equation}
|\psi_k(t)\rangle=\frac{\big[\cos(\Omega_kt)-i\,\sin(\Omega_kt)\,(\mathcal A_k\sigma^z-B_k\sigma^y)/\Omega_k\big]|\psi_k(0)\rangle}{\big\|\big[\cos(\Omega_kt)-i\,\sin(\Omega_kt)\,(\mathcal A_k\sigma^z-B_k\sigma^y)/\Omega_k\big]|\psi_k(0)\rangle\big\|},
\end{equation}
where the scalar $-i\gamma/2$ has dropped out after normalization. Differentiating $\Omega_k^2$, one finds $\partial_k\Omega_k=J\sin k\,(4h-i\gamma)/\Omega_k$, so that $v^{\rm nc}_k$ in Eq.~\eqref{eq:tfim_velocity} reduces to $v_{\rm I}(k)$ at $\gamma=0$.
 
\paragraph{Occupations.---}
Let $|\phi_k^+\rangle$ be the eigenvector of $\mathsf h_k$ with eigenvalue $+\epsilon_{\rm I}(k)$, in which both Bogoliubov modes $\eta_{\pm k}$ are occupied. The occupation entering Eq.~\eqref{eq:tfim_nc} is then
\begin{equation}
\vartheta_k^{\rm nc}(t)=\big|\langle\phi_k^+|\psi_k(t)\rangle\big|^2 .
\label{eq:supp_theta}
\end{equation}
This occupation is conserved at $\gamma=0$, where Eq.~\eqref{eq:tfim_nc} reduces to the unitary quasiparticle formula~\cite{Fag2008,Iannotti2026}. At long times, instead, $|\psi_k(t)\rangle$ converges to the right eigenvector of $\mathcal A_k\sigma^z-B_k\sigma^y$ with eigenvalue $s_k\Omega_k$, where $s_k=\mathrm{sgn}\,\mathrm{Im}\,\Omega_k$, i.e., to the pair state of the non-Hermitian vacuum~\cite{Zer23,Tur22}.
 
\begin{figure}[htbp]
\centering
\includegraphics[width=1.0\linewidth]{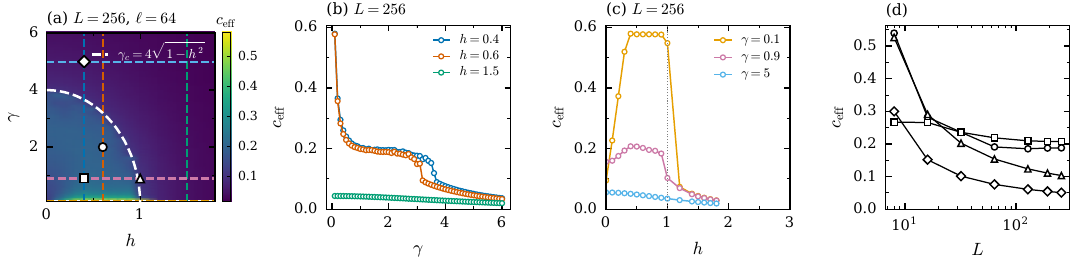}
\caption{Stationary FNLM of the no-click dynamics of the monitored Ising chain, obtained from the non-Hermitian vacuum ($J=1$). (a) Effective coefficient $c_{\rm eff}\equiv\Mf^{\rm st}_\ell/\ln\ell$ in the $(h,\gamma)$ plane for $L=256$ and $\ell=64$. Dashed lines mark the cuts shown in (b) and (c), with matching colors, and symbols mark the points shown in (d). (b) $c_{\rm eff}$ against $\gamma$ at $h=0.4$, $0.6$, and $1.5$, and (c) against $h$ at $\gamma=0.1$, $0.9$, and $5$; the dotted line in (c) marks $h=J$. (d) $c_{\rm eff}$ against $L$ at the points marked in (a). Stationary values are averaged over the last $10\%$ of the time window. Large values at weak monitoring and suppressed values at strong monitoring are consistent with the transition from logarithmic to area-law scaling of the entanglement of the non-Hermitian vacuum~\cite{Tur21c,Zer23}.}
\label{fig:tfim_phase_diagram}
\end{figure}

\paragraph{Stationary scaling and quantum jumps.---}
At long times, the no-click state approaches the vacuum of $H_{\rm eff}$, which lies outside the quasiparticle description: the long-time limit of Eq.~\eqref{eq:tfim_nc} would give a volume law. The FNLM of the vacuum (Fig.~\ref{fig:tfim_phase_diagram}) follows the transition of its entanglement from a logarithmic to an area law~\cite{Tur21c,Zer23}, which Eq.~\eqref{eq:tfim_complex} makes transparent. For $h<J$, ${\rm Im}\,\Omega_k$ changes sign at $\cos k_*=h/J$, where $\Omega_{k_*}^2=4(J^2-h^2)-\gamma^2/4$. For $\gamma<\gamma_c=4\sqrt{J^2-h^2}$, $\Omega_{k_*}$ is real, the vacuum branch $s_k$ switches at $k_*$, and this Fermi-point-like discontinuity yields a logarithmic law; for $\gamma>\gamma_c$, or for $h>J$, the pair state is smooth and the FNLM obeys an area law. Accordingly, $c_{\rm eff}$ is finite inside the line $\gamma_c(h)$ and drops across it, both as a function of $\gamma$ and of $h$ [Figs.~\ref{fig:tfim_phase_diagram}(b) and (c)]; at weak monitoring, it drops at $h\approx J$, where $\gamma_c$ vanishes, and decreases for $h\to0$, where the Ising dispersion becomes flat. With increasing $L$, $c_{\rm eff}$ converges to a finite value inside the line and decreases outside it [Fig.~\ref{fig:tfim_phase_diagram}(d)].

\begin{figure}[htbp]
\centering
\includegraphics[width=1.0\linewidth]{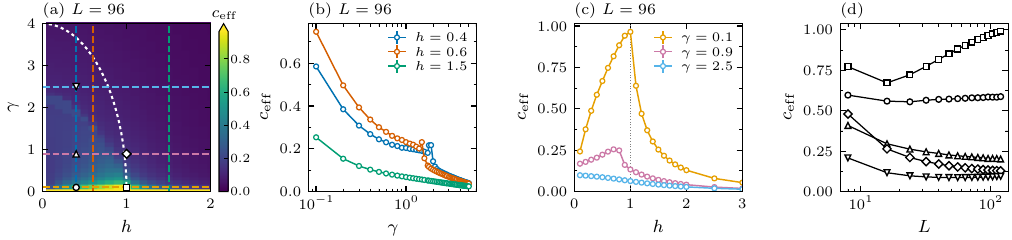}
\caption{Stationary FNLM in the quantum-jump dynamics of the Ising chain with physical occupation monitoring ($J=1$, open chains, $\ell=L/4$ adjacent to the boundary, $N_{\rm traj}=500$). (a) Effective coefficient $c_{\rm eff}\equiv\avg{\Mf}^{\rm st}_\ell/\ln\ell$ in the $(h,\gamma)$ plane at $L=96$. Dashed lines mark the cuts shown in (b) and (c), with matching colors, and symbols mark the points shown in (d). (b) $c_{\rm eff}$ against $\gamma$ at $h=0.4$, $0.6$, and $1.5$, and (c) against $h$ at $\gamma=0.1$, $0.9$, and $5$, both at $L=96$; the dotted line in (c) marks $h=J$. (d) $c_{\rm eff}$ against $L$ at the points marked in (a), with the same symbols. At weak monitoring, $c_{\rm eff}$ is sharply peaked at $h\approx J$, where it grows with $L$. Stationary values are averaged over the last $10\%$ of the time window.}
\label{fig:ceff}
\end{figure}

\begin{figure}[htbp]
\centering
\includegraphics[width=1.0\linewidth]{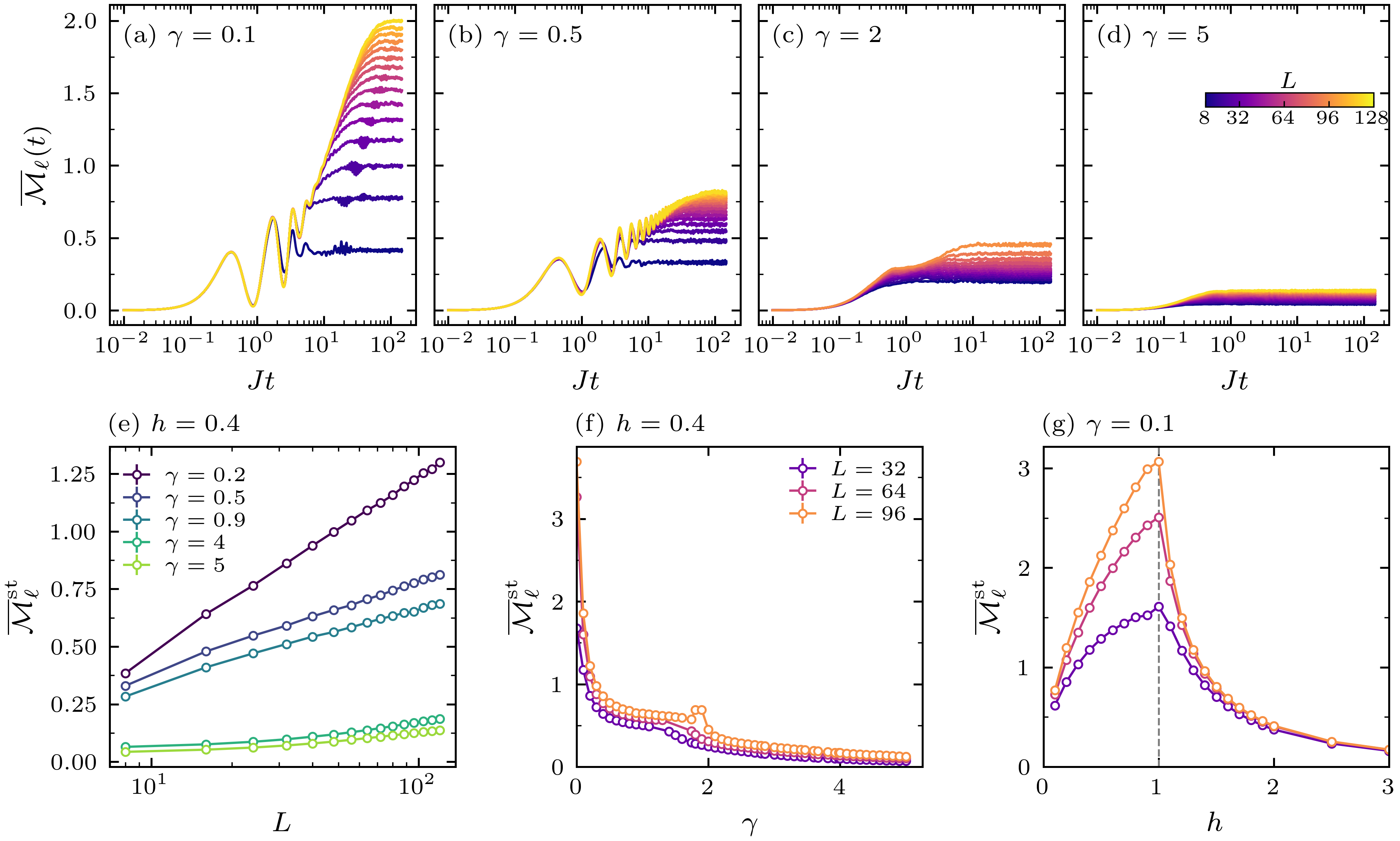}
\caption{Dynamics and size dependence of the FNLM in the quantum-jump dynamics of the Ising chain with physical occupation monitoring (parameters as in Fig.~\ref{fig:ceff}). (a)--(d) Trajectory-averaged FNLM $\avg{\Mf}_\ell(t)$ against $t$ for $\gamma=0.1,0.5,2,5$ at $h=0.4$, for $L=8,\dots,128$ (color scale). (e) Stationary value $\avg{\Mf}^{\rm st}_\ell$ against $L$ at $h=0.4$ for $\gamma=0.2,0.5,0.9,4,5$: straight lines on the logarithmic axis correspond to $\avg{\Mf}^{\rm st}_\ell\propto\ln\ell$. (f) $\avg{\Mf}^{\rm st}_\ell$ against $\gamma$ at $h=0.4$, and (g) against $h$ at $\gamma=0.1$, for $L=32,64,96$; the dashed line in (g) marks $h=J$. Standard errors are shown as shaded bands or error bars.}
\label{fig:ceff_Lscaling}
\end{figure}

 For quantum jumps, the collapse ansatz does not apply in its current form because the Ising Hamiltonian contains pairing terms, so fermion number is not conserved (only fermion parity is); we therefore characterize this regime numerically. The map of $c_{\rm eff}$ (Fig.~\ref{fig:ceff}) has the same structure as in the no-click limit: $c_{\rm eff}$ is large at weak monitoring for $h<J$ and drops across a boundary, at $h\approx J$ for $\gamma=0.1$ and at $\gamma\approx2$ for $h=0.4$ and $0.6$, i.e., at weaker monitoring than $\gamma_c$; correspondingly, the slope of $\avg{\Mf}^{\rm st}_\ell$ against $\ln L$ drops between $\gamma=0.9$ and $\gamma=4$ [Fig.~\ref{fig:ceff_Lscaling}(e)]. The accessible sizes, $L\leq128$, are however smaller than in the no-click limit, and finite-size effects are important. First, $c_{\rm eff}$ can decrease with $L$ due to strong finite size effects, as at $(h,\gamma)=(0.4,0.9)$ [Figs.~\ref{fig:ceff}(d) and~\ref{fig:ceff_Lscaling}(e)]. Second, at $(1,0.1)$, $c_{\rm eff}$ still grows with $L$, so that the FNLM grows faster than $\ln\ell$ near $h=J$. Third, the boundary is broader than in the no-click limit, and its position may drift with $L$. For monitored Ising chains, field theory predicts $\ln^2\ell$ entanglement at weak monitoring~\cite{Fava2023,gal2025replicafieldtheory}; for the FNLM, no prediction is available, and we do not attempt to identify its asymptotic scaling or to locate a transition.

\section{Numerical methods}
\label{sec:supp_numerics}
 
\paragraph{Number-conserving chains.---}
Since the conditional states remain Slater determinants of $N$ particles, we represent each trajectory by an $L\times N$ matrix $B$ whose orthonormal columns are the occupied orbitals in momentum space, from which all correlations follow after a Fourier transformation to real space. For the period-$p$ initial state, the columns are the plane-wave representations of the occupied sites $j=0,p,2p,\dots$. Between jumps, the normalized evolution is unitary, and each orbital simply acquires the phases $e^{-i\epsilon_kt}$, which we apply exactly. Moreover, since the total jump rate $R=\gamma N$ does not depend on the state, the waiting times between consecutive jumps are exponentially distributed with mean $1/(\gamma N)$ and can be sampled directly~\cite{DalibardCastinMolmer1992,Daley2014}. The simulation is therefore exact, with no time discretization. At each jump, the site $j$ is drawn with probability $\langle n_j\rangle/N$, computed from the real-space orbitals obtained by a fast Fourier transform, and the update of Eq.~\eqref{eq:supp_jump_map} is implemented at the level of the orbitals: a Householder reflection rotates the occupied orbitals such that the first one becomes parallel to the projection of $|j\rangle$ onto the occupied subspace, this orbital is replaced by $|j\rangle$, and the remaining $N-1$ orbitals, which are orthogonal to $|j\rangle$, are kept. At each measurement time, the reduced correlation matrix $C_A$ is obtained from the first $\ell$ rows of the real-space orbitals, and its eigenvalues give both the entanglement entropy and the FNLM, Eq.~\eqref{eq:magic}. The interval $A$ consists of $\ell$ consecutive sites of a ring of $L$ sites. For the inhomogeneous quenches, we use open chains and represent the occupied orbitals in real space; between jumps they are evolved exactly through the sine transform of Sec.~\ref{sec:supp_models}, and the jump update is the same. Trajectory averages and standard errors are computed over $N_{\rm traj}=200$ independent trajectories for all the data in Figs.~\ref{fig:four_models_magic}, \ref{fig:scaling_analysis}, \ref{fig:junction}, \ref{fig:periods_g01}, and \ref{fig:periods_g005}, and over $N_{\rm traj}=100$ trajectories for Fig.~\ref{fig:domain_wall}. For the other periods (Sec.~\ref{sec:supp_periods}), we use a ring of $L=420$ sites, which is divisible by $p=3,\dots,7$, times up to $t=6/\gamma$, and the subsystem sizes given in the captions; the weights in Table~\ref{tab:weights} are extracted from $N_{\rm traj}=100$ trajectories with the same time window, using $L=420$ for $p=3,\dots,7$ and $L=360$ for $p=8,9$. Note that the overall sign of the hopping amplitude is immaterial: since $H$, the jump operators, and the initial state are real, the trajectories generated by $-H$ are the complex conjugates of those generated by $H$, with the same jump statistics, and $C_A^*$ has the same spectrum as $C_A$.

\paragraph{Theory curves.---}
The momentum integrals are evaluated on $N_k=1000$ points, and the $x$-integral in Eq.~\eqref{eq:multiplet_magic} is computed exactly, since $m(k,x,t)$ is piecewise constant in $x$. For the renewal term, we use the identity
\begin{equation}
e^{-\gamma t}\min(2|v_k|t,\ell)+\gamma\int_0^t\!da\,e^{-\gamma a}\min(2|v_k|a,\ell)=\frac{2|v_k|}{\gamma}\Big(1-e^{-\gamma\min(t,\,\ell/2|v_k|)}\Big),
\label{eq:supp_identity}
\end{equation}
which is the pair-only collapse prediction of Ref.~\cite{Cao2019}. Equation~\eqref{eq:hybrid} is then evaluated as the pair-only prediction with weight $\ms(\rho)$, plus $e^{-\gamma t}$ times the difference between the multiplet prediction and the unitary pair prediction with the same weight $\ms(\rho)$, which is algebraically identical. All theory curves use the infinite-line geometry; see Sec.~\ref{sec:supp_stationary} for the ring correction.

For the domain wall, Eq.~\eqref{eq:supp_jump_ghd} is integrated with a semi-Lagrangian scheme (free streaming along characteristics followed by exact relaxation over each time step), with $\Delta x=0.5$, $\Delta t=0.25$ and 256 momenta, and the integrals of Eq.~\eqref{eq:renewal} are evaluated on 400 momenta and 400 ages.
 
\paragraph{Ising chain.---}
For the quantum-jump dynamics, since the fermion number is not conserved, we represent the conditional state by the $2L\times2L$ Nambu correlation matrix $G$, with normal and anomalous blocks $C$ and $F$, on an open chain with $J=1$. The initial state has $C=F=0$, i.e., it is the fermionic vacuum, which corresponds to the fully polarized product state $|\uparrow\uparrow\cdots\rangle$. Between jumps, the normalized no-click evolution obeys a closed equation for $G$, which is nonlinear because of the normalization. We integrate it with a fourth-order Runge--Kutta scheme with time step $dt=10^{-2}$, re-imposing the Hermiticity of $C$ and the antisymmetry of $F$ after each step. Within each step, a jump occurs with probability $1-e^{-R\,dt}$, where $R=\gamma\sum_j\langle L_j^\dagger L_j\rangle$ is evaluated at the beginning of the step, and its site is chosen with probability proportional to the local rate. The jump is then applied through the Gaussian update of $C$ and $F$, which generalizes Eq.~\eqref{eq:supp_jump_map} to states without fixed particle number. At most one jump is applied per step, so the scheme requires $R\,dt\ll1$. The entanglement entropy and the FNLM are computed from the spectrum of the restriction of $G$ to $A=\{1,\dots,\ell\}$, with $\ell=L/4$ adjacent to the open boundary. For each $(L,h,\gamma)$, we average over $N_{\rm traj}=500$ trajectories up to $t=e^5\approx148$, with independent random streams generated from a fixed seed. We simulate $L=8,\dots,128$ with $\ell=L/4$, and average the FNLM over the last $10\%$ of the time window to obtain its stationary value $\avg{\Mf}^{\rm st}_\ell$ and the effective coefficient $c_{\rm eff}\equiv\avg{\Mf}^{\rm st}_\ell/\ln\ell$. The map in Fig.~\ref{fig:ceff}(a) is computed at $L=96$; the cuts and the size dependence are shown in Figs.~\ref{fig:ceff}(b)--(d) and~\ref{fig:ceff_Lscaling}.

\paragraph{No-click dynamics in momentum space.---}
Starting from the fermionic vacuum, the normalized no-click state remains translation invariant and couples only the modes $k$ and $-k$. It is therefore fully described by the $2\times2$ Nambu blocks
\begin{equation}
\mathcal G_k=\begin{pmatrix}\langle c_kc_k^\dagger\rangle&\langle c_kc_{-k}\rangle\\ \langle c_{-k}^\dagger c_k^\dagger\rangle&\langle c_{-k}^\dagger c_{-k}\rangle\end{pmatrix}\equiv\begin{pmatrix}n_k&\phi_k\\ \phi_k^*&d_k\end{pmatrix},\qquad k\in(0,\pi),
\end{equation}
with $\mathcal G_k(0)={\rm diag}(1,0)$. Each block obeys the normalized no-click equation
\begin{equation}
\partial_t\mathcal G_k=2i[\mathsf H_k,\mathcal G_k]-\gamma\,\mathcal Q(\mathcal G_k),\qquad
\mathsf H_k=\begin{pmatrix}h-J\cos k&-iJ\sin k\\ iJ\sin k&-(h-J\cos k)\end{pmatrix},
\end{equation}
where the nonlinear term, which stems from the normalization, reads
\begin{equation}
\mathcal Q(\mathcal G_k)=\begin{pmatrix}n_k^2-|\phi_k|^2-n_k&\phi_k(n_k-d_k)\\ \phi_k^*(n_k-d_k)&|\phi_k|^2-d_k^2+d_k\end{pmatrix}.
\end{equation}
We integrate these equations with a fourth-order Runge--Kutta scheme, on the grid $k_m=(m+\frac12)\pi/N_k$ with $N_k=4000$, re-imposing the Hermiticity of $\mathcal G_k$ after each step. Since the blocks are decoupled, this describes the infinite chain directly, with no finite-size effects. The real-space correlations depend only on $r=x-y$,
\begin{align}
C(r)&=\langle c_xc_y^\dagger\rangle=\int_0^\pi\frac{dk}{2\pi}\Big[e^{ikr}n_k+e^{-ikr}(1-d_k)\Big],\\
F(r)&=\langle c_xc_y\rangle=\int_0^\pi\frac{dk}{2\pi}\big(e^{ikr}-e^{-ikr}\big)\phi_k,
\end{align}
so that the restriction to an interval of $\ell$ sites is given by the Toeplitz matrices $(C_A)_{xy}=C(x-y)$ and $(F_A)_{xy}=F(x-y)$, Hermitian and antisymmetric, respectively. The FNLM then follows from the spectrum of the $2\ell\times2\ell$ Nambu matrix built from $C_A$ and $F_A$, as in Eq.~\eqref{eq:magic}.


\end{document}